\documentclass[letterpaper,11pt]{article}

\usepackage{jheppub}

\usepackage{jheppub}
\usepackage{subfig}
\usepackage{xspace}

\usepackage{xcolor}

\newcommand{\fjcontrib}{\textsc{FastJet Contrib}}

\newcommand{\SCETa}{\mbox{${\rm SCET}_{\rm I}$}\xspace}
\newcommand{\SCETb}{\mbox{${\rm SCET}_{\rm II}$}\xspace}

\newcommand{\eq}[1]{eq.~\eqref{eq:#1}}

\newcommand{\abs}[1]{\lvert#1\rvert}

\newcommand{\df}{\mathrm{d}}

\newcommand{\nn}{\nonumber}

\newcommand{\Ecm}{E_\mathrm{cm}}

\newcommand{\Tau}{\mathcal{T}}

\newcommand{\GeV}{\text{GeV}}

\allowdisplaybreaks[4]

\DeclareRobustCommand{\Sec}[1]{section~\ref{#1}}

\DeclareRobustCommand{\Tab}[1]{table~\ref{#1}}
\DeclareRobustCommand{\Tabs}[2]{tables~\ref{#1} and \ref{#2}}
\DeclareRobustCommand{\Fig}[1]{figure~\ref{#1}}
\DeclareRobustCommand{\Figs}[2]{figures~\ref{#1} and \ref{#2}}
\DeclareRobustCommand{\Eq}[1]{eq.~(\ref{#1})}

\DeclareRobustCommand{\Ref}[1]{ref.~\cite{#1}}
\DeclareRobustCommand{\Refs}[1]{refs.~\cite{#1}}

\newcommand{\be}{\begin{equation}}
\newcommand{\ee}{\end{equation}}

\newcommand{\jet}{{\text{jet}}}
\newcommand{\beam}{{\text{beam}}}

\preprint{ 
\begin{flushright}
MIT--CTP 4674\\
DESY 15-087
\end{flushright}}
 
\title{XCone:  N-jettiness as an Exclusive Cone Jet Algorithm}

\author[a]{Iain~W.~Stewart,}
\author[b]{Frank~J.~Tackmann,} 
\author[a]{Jesse~Thaler,}
\author[c]{\\ Christopher K. Vermilion,}
\author[a]{and Thomas~F.~Wilkason}

\affiliation[a]{Center for Theoretical Physics, Massachusetts Institute of
  Technology, Cambridge, MA 02139, USA}
\affiliation[b]{Theory Group, Deutsches Elektronen-Synchrotron (DESY), D-22607 Hamburg, Germany}
\affiliation[c]{Ernest Orlando Lawrence Berkeley National Laboratory, University of California, Berkeley, CA 94720, USA}

\emailAdd{iains@mit.edu}
\emailAdd{frank.tackmann@desy.de}
\emailAdd{jthaler@mit.edu}
\emailAdd{christopher.vermilion@gmail.com}
\emailAdd{tjwilk@mit.edu}

\abstract{
We introduce a new jet algorithm called XCone, for eXclusive Cone, which is based on minimizing the event shape $N$-jettiness.  Because $N$-jettiness partitions every event into $N$ jet regions and a beam region, XCone is an exclusive jet algorithm that always returns a fixed number of jets. We use a new ``conical geometric'' measure for which well-separated jets are bounded by circles of radius $R$ in the rapidity-azimuth plane, while overlapping jet regions automatically form nearest-neighbor ``clover jets''. This avoids the split/merge criteria needed in inclusive cone algorithms. A key feature of XCone is that it smoothly transitions between the resolved regime where the $N$ signal jets of interest are well separated and the boosted regime where they overlap. The returned value of $N$-jettiness also provides a quality criterion of how $N$-jet-like the event looks.  We also discuss the $N$-jettiness factorization theorems that occur for various jet measures, which can be used to compute the associated exclusive $N$-jet cross sections. In a companion paper~\cite{Thaler:2015xaa}, the physics potential of XCone is demonstrated using the examples of dijet resonances, Higgs decays to bottom quarks, and all-hadronic top pairs.
}

\begin{document}

\maketitle

\section{Introduction}
\label{sec:intro}

Collisions at the Large Hadron Collider (LHC) are dominated by jets, collimated sprays of hadrons arising from the fragmentation of energetic quarks and gluons.  Jets are crucial to connect the observed hadronic final state to the short-distance hard interaction.  Fundamentally, the definition of a hadronic jet is ambiguous, since there is no unique way to map color-singlet hadrons to color-carrying partons.  Moreover, different physics applications can benefit from different jet definitions.  For these reasons, a wide variety of jet algorithms have been proposed to identify and study jets  \cite{Ellis:2007ib,Salam:2009jx}, though currently, most LHC measurements involve jets clustered with the anti-$k_T$ algorithm \cite{Cacciari:2008gp}.

In this paper, we present a new jet algorithm that we call ``XCone''. It is based on minimizing the event shape $N$-jettiness~\cite{Stewart:2010tn} and uses developments from the jet shape $N$-subjettiness~\cite{Thaler:2010tr, Thaler:2011gf}.  The key feature is that $N$-jettiness defines an \emph{exclusive cone} jet algorithm.  Like the exclusive $k_T$ algorithm \cite{Catani:1993hr}, our XCone algorithm returns a fixed number of jets, relevant for physics applications where the number of jets is known in advance.  Like anti-$k_T$ jets~\cite{Cacciari:2008gp}, XCone jets are nearly conical for well-separated jets, such that they have fixed active jet areas \cite{Cacciari:2007fd,Cacciari:2008gn}. Typically, when using other jet algorithms, the boosted regime of overlapping jets requires separate analysis strategies using fat jets with substructure~\cite{Abdesselam:2010pt, Altheimer:2012mn, Altheimer:2013yza, Adams:2015hiv}. In contrast, with XCone the jets remain resolved even when jets are overlapping in the boosted regime. In this way, XCone smoothly interpolates between the resolved regime of widely-separated jets and the boosted regime of collimated subjets. This feature will be explored in more depth in a companion paper~\cite{Thaler:2015xaa}, which demonstrates the application of XCone for the examples of dijet resonances, Higgs decays to bottom quarks, and all-hadronic top pairs.

The possibility of using $N$-jettiness as a jet algorithm was already pointed out in \Ref{Stewart:2010tn} and further explored in \Ref{Thaler:2011gf}. Here, we more fully develop the idea of $N$-jettiness jets and present a concrete implementation of the XCone algorithm. As a global event shape, $N$-jettiness measures the degree to which the hadrons in the final state are aligned along $N$ jet axes or the beam direction.  It was originally introduced to veto additional jets in an event, providing a way to define and resum exclusive $N$-jet cross sections~\cite{Stewart:2009yx, Stewart:2010tn, Jouttenus:2011wh}.\footnote{The reader should be aware that there are two different definitions of ``exclusive'' which are both standard in their respective contexts.  An exclusive $N$-jet \emph{algorithm} is one that returns exactly $N$ jets, regardless of what happens in the rest of the event.  An exclusive $N$-jet \emph{cross section} is the rate to produce exactly $N$ jets, with a restriction on what happens in the rest of the event.  XCone is an exclusive $N$-jet algorithm, but it can be used either to measure inclusive $N$-jet cross sections (if there are no restrictions made on unclustered particles) or an exclusive $N$-jet cross section (if there is a restriction, say, that $\Tau_N < \Tau_{\rm cut}$).}  $N$-jettiness was later adapted to the jet shape $N$-subjettiness \cite{Thaler:2010tr}, which is an efficient measure to identify $N$-prong boosted hadronic objects such as top quarks, $W$/$Z$ bosons, and Higgs bosons within a larger jet (see also \cite{Kim:2010uj}).  By minimizing $N$-(sub)jettiness, one can directly identify $N$ (sub)jet directions, and a fast algorithm to perform this minimization was presented in \Ref{Thaler:2011gf}.
$N$-jettiness jets have been used to resum the invariant mass of nearby jets~\cite{Bauer:2011uc}, to make predictions for jet mass spectra~\cite{Jouttenus:2013hs, Stewart:2014nna}, for studying DIS and nuclear dynamics~\cite{Kang:2013nha, Kang:2013lga, Kang:2012zr, Kang:2013wca, Kang:2014qba, Kang:2015dga}, and to define recoil-free jet observables \cite{Larkoski:2014uqa}. As an $N$-jet resolution variable, $N$-jettiness has been utilized to combine perturbative calculations with parton showers in {\sc Geneva}~\cite{Alioli:2012fc}, and very recently to define a powerful subtraction scheme for fixed-order calculations at next-to-next-to-leading order~\cite{Boughezal:2015dva, Gaunt:2015pea}.

As we will see, there is considerable flexibility in precisely how one defines $N$-jettiness, and several different $N$-jettiness measures yielding different jet regions have been considered before~\cite{Stewart:2010tn, Jouttenus:2011wh, Jouttenus:2013hs, Thaler:2010tr, Thaler:2011gf}.  Here, as the XCone default, we propose a ``conical geometric'' measure that incorporates the insights from the different previous use cases.  This measure is based on the dot product between particles and lightlike axes as in \Ref{Stewart:2010tn} but incorporates an angular exponent $\beta$ as in \Ref{Thaler:2011gf}, as well as a beam exponent $\gamma$ for additional flexibility (see \Tab{tab:measures} below).  Crucially for the purposes of jet finding at the LHC, this measure yields (nearly) conical jets over a wide rapidity range, and the user can choose the desired jet radius $R$.

For most physics applications, we propose a default setting of $\beta = 2$ and $\gamma = 1$, which acts similarly to existing cone algorithms (see e.g.~\cite{Sterman:1977wj, Blazey:2000qt,Ellis:2001aa,Salam:2007xv}) in that the resulting jet regions are (approximately) stable cones where the jet momenta and the jet axes align.  The key difference to algorithms like SISCone \cite{Salam:2007xv} is that XCone does not require a split/merge step.  In particular, typical inclusive cone algorithms have an overlap parameter which determines whether two abutting stable cones should be joined or remain separate.  By contrast, XCone only requires setting the jet radius $R$ and the number of desired jets $N$, and the split/merge decision is determined dynamically through $N$-jettiness minimization.  In a companion paper \cite{Thaler:2015xaa}, we show examples of quasi-boosted kinematics that capitalize on this exclusive approach to cone jet finding.

There are interesting connections between $N$-jettiness minimization and previous work to define jets via cluster optimization \cite{Ellis:2001aa,Berger:2002jt,Angelini:2002et,Angelini:2004ac,Grigoriev:2003yc,Grigoriev:2003tn,Chekanov:2005cq,Lai:2008zp,Volobouev:2009rv,Mackey:2015hwa}.  Stable cone finding is closely related to 1-jettiness minimization with $\beta = 2$ \cite{Ellis:2001aa}, and similar algorithms are relevant for a recently proposed ``jet function''\footnote{The name jet function in this context should not be confused with the more standard usage in the context of factorization of cross sections into hard, soft, and jet functions, e.g.~\cite{Sterman:1986aj,Catani:1989ne,Korchemsky:1994jb}. Here our primary use of the name jet function will be in this factorization context, see sec.~\ref{sec:factorization}.  } optimization strategy \cite{Georgi:2014zwa,Ge:2014ova,Bai:2014qca}. One can even prove an exact equivalence between these algorithms when finding a single cone jet of fixed opening angle~\cite{Thaler:2015uja}. Finding the thrust axis \cite{Farhi:1977sg} is related to $2$-jettiness minimization with $\beta = 2$.\footnote{Naively, one might think that spherocity \cite{Georgi:1977sf} should be related to $2$-jettiness with $\beta = 1$.  However, minimizing this quantity does not give rise to the spherocity axis, but rather to kinked broadening axes \cite{Larkoski:2014uqa}.}   There is also an observable called triplicity \cite{Brandt:1978zm} which is related to $3$-jettiness.  For a general $N$, $k$-means clustering \cite{Lloyd82leastsquares} (with $k = N$) is a type of $N$-jettiness minimization, with $\beta = 2$ corresponding to traditional $k$-means and $\beta = 1$ corresponding to $R1$-$k$-means \cite{Ding:2006:RPR:1143844.1143880}.  In all these cases, $N$-jettiness minimization is an infrared and collinear (IRC) safe procedure.

Because cluster optimization is a difficult computational problem, our practical XCone implementation will use recursive clustering algorithms \cite{Catani:1993hr, Ellis:1993tq, Dokshitzer:1997in, Wobisch:1998wt, Wobisch:2000dk} to approximate $N$-jettiness minima.  Roughly speaking, we run a generalized $k_T$ clustering algorithm to determine IRC-safe seed jet axes as a starting point for an iterative one-pass minimization algorithm, in which $N$-jettiness is used to find the final jet axes and define the jet regions.  Separating jet axes finding from jet region finding appeared previously in the context of recoil-free jets \cite{Larkoski:2014uqa,Larkoski:2014bia}, where a fixed radius cone was centered on winner-take-all axes \cite{Bertolini:2013iqa,Larkoski:2014uqa,Salambroadening} or broadening axes \cite{Thaler:2011gf,Larkoski:2014uqa}.  XCone allows us to extend this strategy to $N$-jet events, with $\beta = 1$ yielding recoil-free jets and $\beta = 2$ yielding traditional cones where the jet axes and jet momenta are (nearly) aligned.

A key feature of the measures we consider, including the default XCone measure, is that $N$-jettiness can be decomposed into a direct sum of contributions from the jet and beam regions.  When utilizing measures with this property, there exist active-parton factorization theorems for $N$-jettiness cross sections valid to all orders in $\alpha_s$.  Furthermore, the default XCone measure is linear in the particle momenta which greatly simplifies the calculation of the perturbative jet and soft functions needed to determine the $N$-jettiness cross section.  Thus, the ingredients needed for higher-order logarithmic resummation or fixed-order calculations are simpler for jets defined with the XCone algorithm, in contrast for example to those defined with clustering algorithms like anti-$k_T$.  We will discuss these factorization theorems in some detail for various choices of $N$-jettiness measures, including the XCone default.

The remainder of this paper is organized as follows.  In \Sec{sec:algorithm}, we review how to define an exclusive jet algorithm via minimizing $N$-jettiness.  We then discuss a variety of $N$-jettiness measures in \Sec{sec:measure}, including the conical geometric measure that is the basis for XCone.  In \Sec{sec:details}, we discuss some details of our XCone implementation, in particular the choice of seed axes for finding a (local) $N$-jettiness minimum.  In \Sec{sec:factorization}, we discuss the factorization theorems for $N$-jettiness with various measures.  This section is more theoretically technical than the others and may be skipped by readers not interested in this factorization. We conclude in \Sec{sec:conclude}. The XCone algorithm is available through the \textsc{Nsubjettiness} \fjcontrib\ \cite{Cacciari:2011ma, fjcontrib} as of version 2.2.0.

\section{\boldmath $N$-jettiness as a Jet Algorithm}
\label{sec:algorithm}

Given a set of normalized lightlike axes $n_A = \{1, \vec{n}_A\}$ with $\vec{n}_A^2 = 1$, $N$-jettiness is defined as\footnote{Here we use a dimension-one definition as in \Refs{Jouttenus:2011wh, Jouttenus:2013hs} instead of the dimensionless $\tau_N$ used in \Ref{Stewart:2010tn}.} 
\be
\label{eq:tauNdef}
\widetilde{\Tau}_N = \sum_i \min\left\{\rho_{\jet}(p_i,n_1), \ldots, \rho_{\jet}(p_i, n_N), \rho_{\beam}(p_i) \right\}
\,.\ee
The sum runs over the four-momenta $p_i$ of all particles that are considered as part of the hadronic final state and should take part in the jet clustering.  The $\rho_\jet(p_i,n_A)$ is a distance measure to the $A$-th axis $n_A$, and $\rho_{\beam}(p_i)$ is a distance measure to the beam. Depending on the context, the beam measure can be separated into two beam regions with lightlike beam axes $n_{a,b}$ and (partonic) center-of-mass rapidity $Y$ such that
\be
\label{eq:rhobeamdivide}
\rho_{\beam}(p_i) \Rightarrow \min \{ \rho_{\beam}(p_i,n_a,Y),  \rho_{\beam}(p_i,n_b,Y) \}
\,.\ee
This form will be relevant for the discussion in \Sec{sec:factorization}.

For a given form of $\rho_\jet$ and $\rho_{\beam}$, the minimum inside $\widetilde{\Tau}_N$ in \eq{tauNdef} partitions the particles $i$ into $N$ jet regions and an unclustered beam region.  To use $N$-jettiness as a jet algorithm, one minimizes $\widetilde{\Tau}_N$ over all possible lightlike axes directions:
\be
\label{eq:mincriteria}
\Tau_N = \min_{n_1, n_2, \ldots, n_N} \widetilde{\Tau}_N.
\ee
The locations of the axes at the minimum define the centers of the jet regions. In previous applications, one uses a separate method to choose the $N$-jettiness axes $n_A$, e.g.\ from the $N$ hardest jets found by some other jet algorithm. One then uses $\widetilde{\Tau}_N$ only for the jet partitioning (in which case there is no need to distinguish $\Tau_N \equiv \widetilde{\Tau}_N$). This use of $\Tau_N$ already provides a well-defined and IRC-safe way to define $N$ exclusive jets. The additional overall minimization in \eq{mincriteria} over the axes $n_A$ promotes $\Tau_N$ to a standalone exclusive jet algorithm.  This axis minimization is nontrivial and we discuss our strategy to perform it in \Sec{sec:details}.\footnote{One might also be able to dynamically determine the total rapidity $Y$ or the beam axes $n_{a,b}$ through minimization, though that feature is currently not present in the XCone code.}  Note that ``minimization'' can refer either to finding the global $\Tau_N$ minimum or using an IRC-safe procedure to find a local $\Tau_N$ minimum, either of which is suitable for the discussion below.

Any choice of measure together with the specific algorithm to minimize $\Tau_N$ defines an exclusive jet algorithm.  In particular, $\Tau_N$ in \Eq{eq:tauNdef} always identifies $N$ jet regions (and one beam region), regardless of how close the axes $n_A$ might be to each other.  When the axes are well separated, the boundary of the jet regions is determined through competition between $\rho_\jet$ and $\rho_\beam$.  When the axes are close together, the jet regions are determined by the competition between different $\rho_\jet$.

\begin{table*}[t]
\small
\begin{center}
\begin{tabular}{ccccc}
\hline \hline
Name & $\rho_{\jet}(p_i,n_A)$ & $\rho_{\beam}(p_i)$ & $A \approx \pi R^2$?\\
\hline \hline
Conical \cite{Thaler:2011gf} &  $p_{Ti}\, \Bigl(\dfrac{R_{iA}}{R}\Bigr)^\beta$ & $p_{Ti}$ & \checkmark \\
General Conical &  $p_{Ti}\,f(p_i)\, \Bigl(\dfrac{R_{iA}}{R}\Bigr)^\beta$ & $p_{Ti}\,f(p_i)$& \checkmark  \\
\hline
Geometric \cite{Jouttenus:2013hs} & $\dfrac{n_A \cdot p_i}{\rho_0}$ & $m_{Ti} e^{-|y_i|}$ \\
Modified Geometric & $\dfrac{n_A \cdot p_i}{\rho_0}$ & $\dfrac{m_{Ti}}{2 \cosh y_i}$ \\
Geometric-$R$ \cite{Jouttenus:2013hs} & $\dfrac{n_A \cdot p_i}{\rho(R, y_A)}$ & $m_{Ti} e^{-|y_i|}$ & \checkmark  \\
Modified Geometric-$R$ & $\dfrac{n_A \cdot p_i}{\rho_C(R,y_A)}$ & $\dfrac{m_{Ti}}{2 \cosh y_i}$ & \checkmark \\
\hline
Conical Geometric & $\dfrac{p_{Ti}}{(2 \cosh y_i)^{\gamma - 1}} \Bigl(\dfrac{2 \, n_A \cdot p_i}{n_{TA} \, p_{Ti}}\,\dfrac{1}{R^2} \Bigr)^{\beta/2}$ & $\dfrac{p_{T i}}{(2 \cosh y_i)^{\gamma - 1}}$  & \checkmark\\
XCone Default ($\beta = 2, \gamma = 1$) & $\dfrac{2 \cosh y_A}{R^2} \, n_A \cdot p_i$ & $p_{T i}$  & \checkmark \\
Recoil-Free Default ($\beta = 1, \gamma = 1$) & $\sqrt{\dfrac{2  \cosh y_A}{R^2}  \, p_{Ti} \,  n_A \cdot p_i}$ & $p_{T i}$  & \checkmark\\
$\beta = 2$, $\gamma = 2$ & $\dfrac{\cosh y_A}{\cosh y_i\, R^2} \, n_A \cdot p_i$ & $\dfrac{p_{T i}}{2 \cosh y_i}$  & \checkmark\\
\hline \hline
\end{tabular}
\end{center}
\caption{$N$-jettiness measures studied in this paper.   The conical geometric measure with $\beta = 2$ and $\gamma =1 $ is the suggested XCone default, giving stable cone jets (like the conical measure) through dot-product distances linear in $p_i$ (like the geometric measures).  The recoil-free variant with $\beta = 1$ centers the jet around its hardest cluster, making the jet regions less sensitive to soft contamination.  In the conical geometric measure, $n_{TA} = 1/\cosh y_A$.   In the (modified) geometric-$R$ measures, $\rho_{(C)}(R,y_A)$ is a rapidity-dependent scale factor that yields jet areas of exactly $\pi R^2$ (though not conical jet boundaries).  The checkmarks indicate measures that yield jets with active areas of $\approx \pi R^2$ for well-separated jets. These active areas are $\pi R^2$ to within $\lesssim 1\%$ over a wide rapidity range (see \Fig{fig:jetarea} below).
}
\label{tab:measures}
\end{table*}

To go from an exclusive jet algorithm to an exclusive \emph{cone} jet algorithm (i.e.~XCone), one wants the jet boundaries to approximate circles in the rapidity-azimuth plane, which can be achieved by an appropriate choice of jet and beam measures.  In \Sec{sec:measure}, we study a variety of jet and beam measures which are summarized in \Tab{tab:measures}. This includes three new measures:  the general conical measure in \Eq{eq:fmeasure} which yields exact cones for widely-separated jets; the modified geometric measure in \Eq{eq:modgeomeasure} whose jet measure is linear in particle momenta like the original geometric measure but exhibits smooth behavior at zero rapidity; and the recommended XCone default in \Eq{eq:xconemeasure} which yields approximate cones and also features this linearity. By construction, the XCone default measure yield jets with approximately fixed active jet areas over a wide range of jet rapidities.

In addition to partitioning the event into jet and beam regions, the returned value of $\Tau_N$ is a quality criterion that measures how well an event is characterized by $N$ jets. The contribution to the $\Tau_N$ value from a given jet provides a measure of how collimated the jet is. For narrow jets (i.e.~small effective jet radius), $\Tau_N$ is typically dominated by the contribution from the beam region.  Thus, for LHC applications, one typically wants $\rho_{\beam}(p_i)$ to be proportional to $p_{Ti}$ (the transverse momentum of particle $i$) such that minimizing $\Tau_N$ results in the least unclustered $p_T$. Larger values of $\Tau_N$, and its beam contribution in particular, then indicate additional activity or hard jets in the event. An improved measure of jet quality can be obtained by examining the individual jet and beam contributions to $\Tau_N$, as in~\cite{Jouttenus:2011wh}:
\begin{align} \label{eq:tauNjetbeam}
\Tau_N= \Tau_N^{\rm beam}  + \Tau_N^{\rm jets} 
    =  \Tau_N^{\rm beam} + \sum_{A=1}^N \Tau_N^{A}  \,.
\end{align}
Here, $\Tau_N^{\rm jets}$ provides a global measure for assessing how collimated the jets are without contamination from the beam region, and one can obtain individual quality measures for each of the $N$ jets by examining their individual numerical contributions $\Tau_N^{A}$ to the total $N$-jettiness.    In \Sec{sec:factorization}, we discuss some of the theoretical aspects involved in calculating $\Tau_N$ as well as the cross section that is fully differential in $\Tau_N^{\rm beam}$ and the $N$ observables $\Tau_N^{A}$.

Before discussing the specific measures, we want to make a general comment about underlying event and pileup, two effects that are known to impact jet reconstruction.  While the \emph{value} of $\Tau_N$ depends strongly on these effects, the jet regions found by minimizing $\Tau_N$ are no more sensitive to underlying event and pileup than traditional jet algorithms.  The reason for this mismatch is that the beam contribution to the $\Tau_N$ value can get large contributions from these effects, but the change in $\Tau_N$ as the axes $n_A$ are varied only depends on hadrons in the vicinity of the jet regions.  This is particularly true for recoil-free measures, where the minimized axis direction is almost entirely insensitive to soft contamination \cite{Larkoski:2014bia}.  For pileup specifically, the minimization in \Eq{eq:mincriteria} remains sensible even with negative energy particles, so one has the option of introducing negative energy ghosts as a way to implement area subtraction~\cite{Cacciari:2007fd, Cacciari:2008gn, Soyez:2012hv}. For isolated jets, one can derive a closed-form integral expression for the active jet area, which depends only mildly on the jet rapidity.

\begin{figure}
\centering
\subfloat[]{\label{fig:jetpicture:a}%
\includegraphics[width=.49\columnwidth]{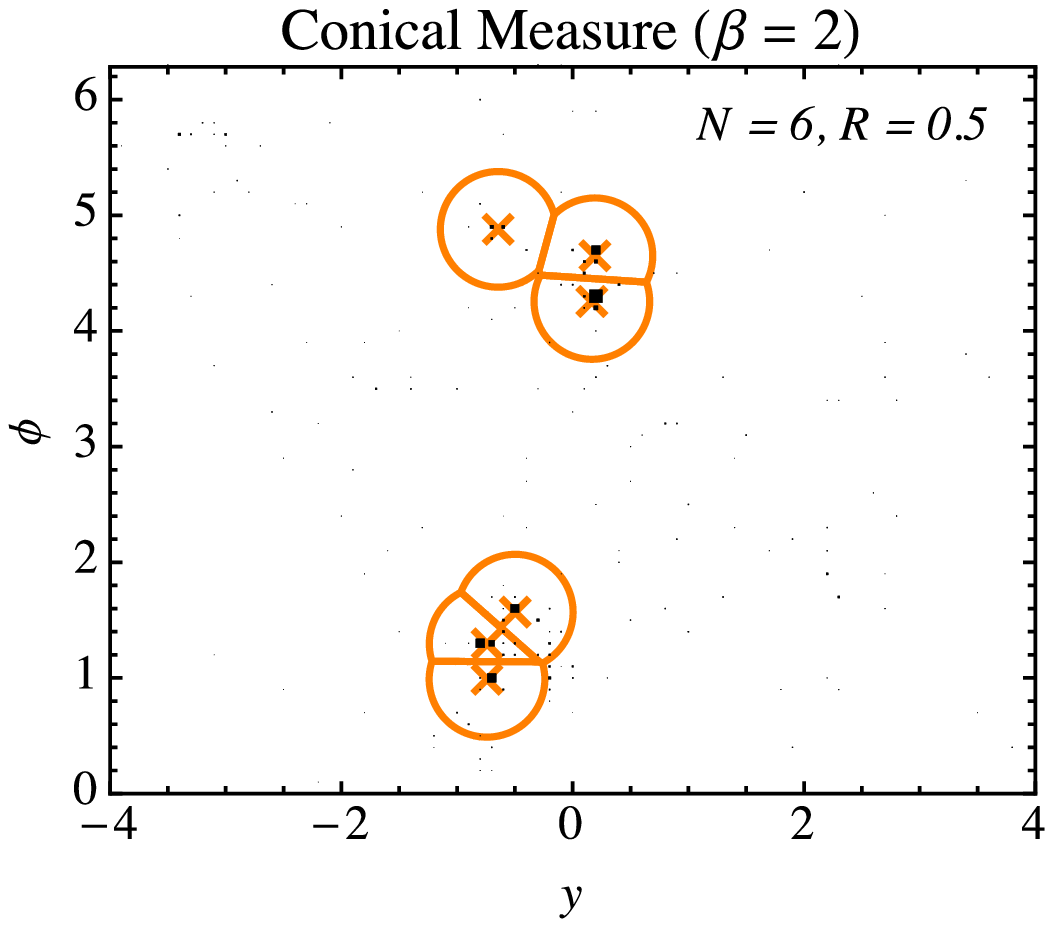}}
\hfill%
\subfloat[]{\label{fig:jetpicture:b}%
\includegraphics[width=.49\columnwidth]{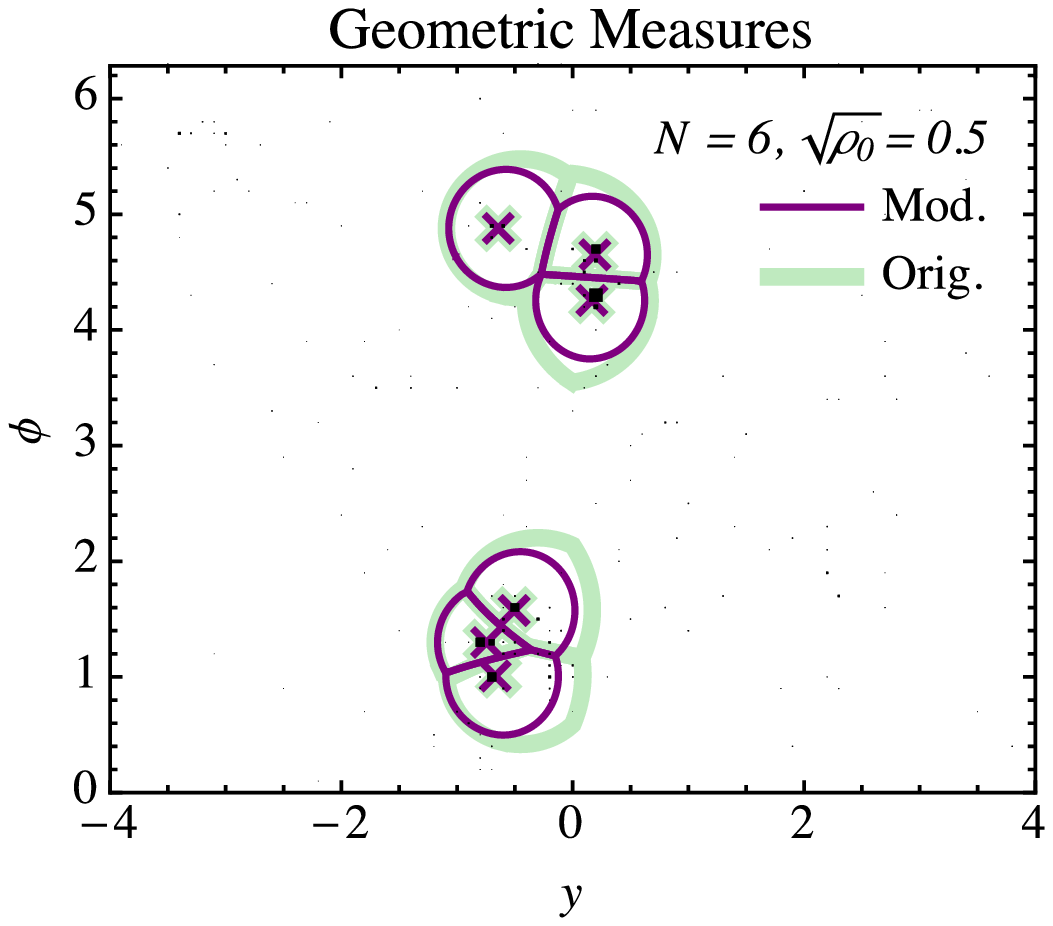}}%
\\
\subfloat[]{\label{fig:jetpicture:c}
\includegraphics[width=.49\columnwidth]{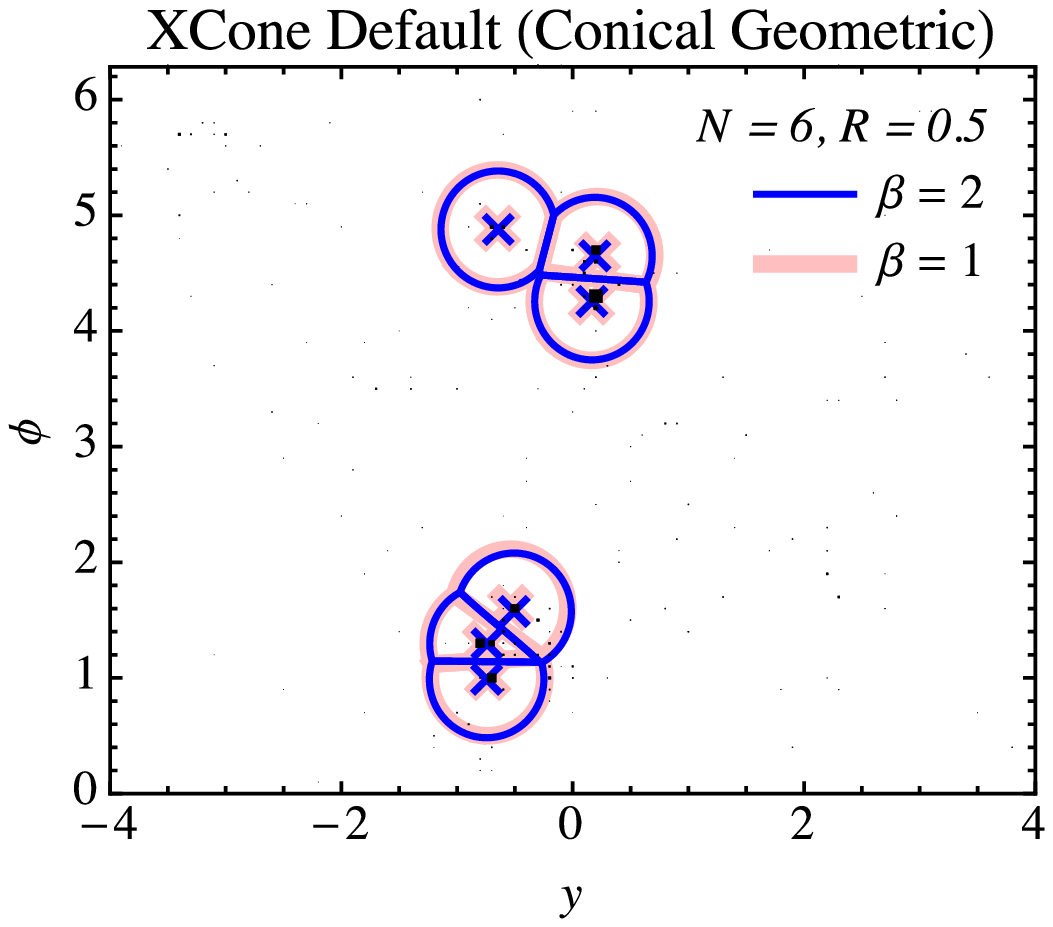}
}
\caption{Jet regions found with various $N$-jettiness measures.  This is a $t\bar{t}$ event from the BOOST 2010 event sample \cite{Abdesselam:2010pt}, and every measure has $N = 6$ and $R = 0.5$.  (a) Conical measure with $\beta = 2$.  (b) Original and modified geometric measures.  (c) Conical geometric measure with $\beta = 2$ (XCone default) and $\beta = 1$ (recoil-free default).  The conical and conical geometric measures yield (approximately) circular jets. For all measures, the overlap region between jets is automatically partitioned by nearest neighbor, as given by the jet measure.}
\label{fig:jetpicture}
\end{figure}

\begin{figure}[p]
\centering
\subfloat[]{\label{fig:jetpicture2:a}%
\includegraphics[width=.49\columnwidth]{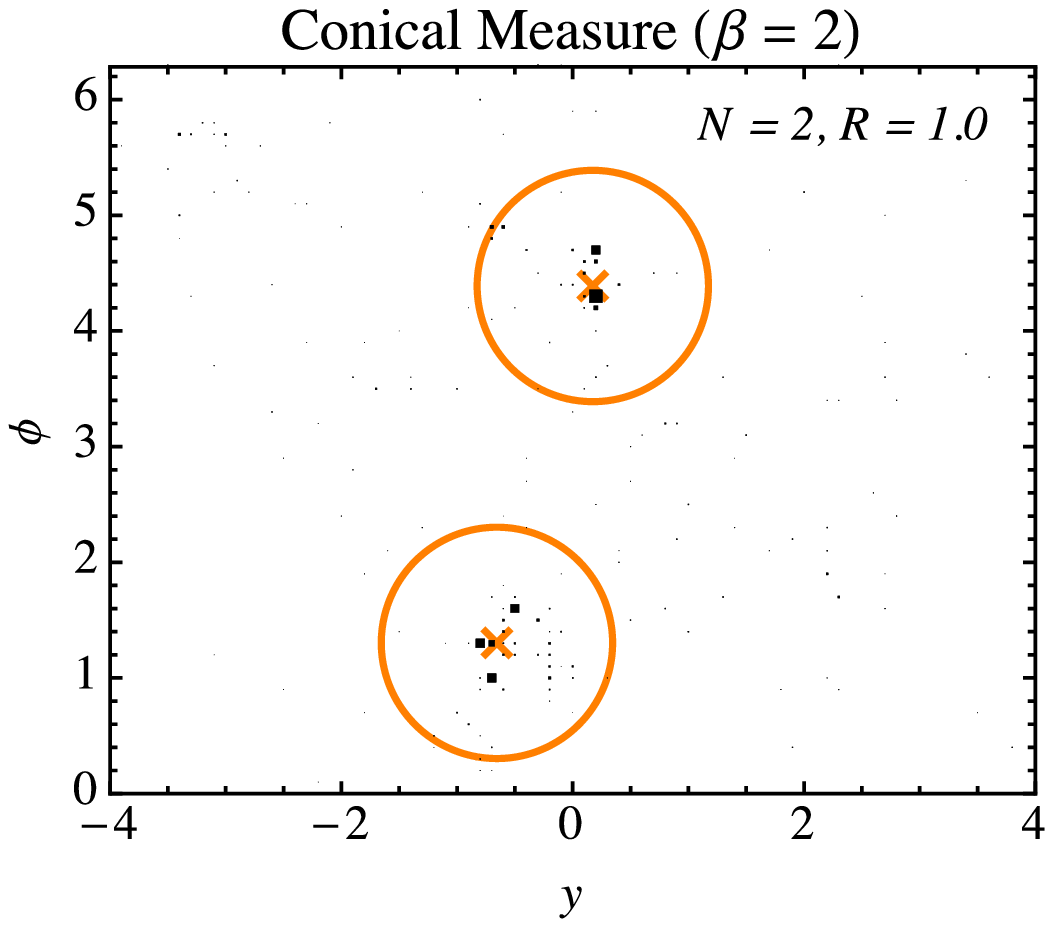}}%
\hfill%
\subfloat[]{\label{fig:jetpicture2:b}
\includegraphics[width=.49\columnwidth]{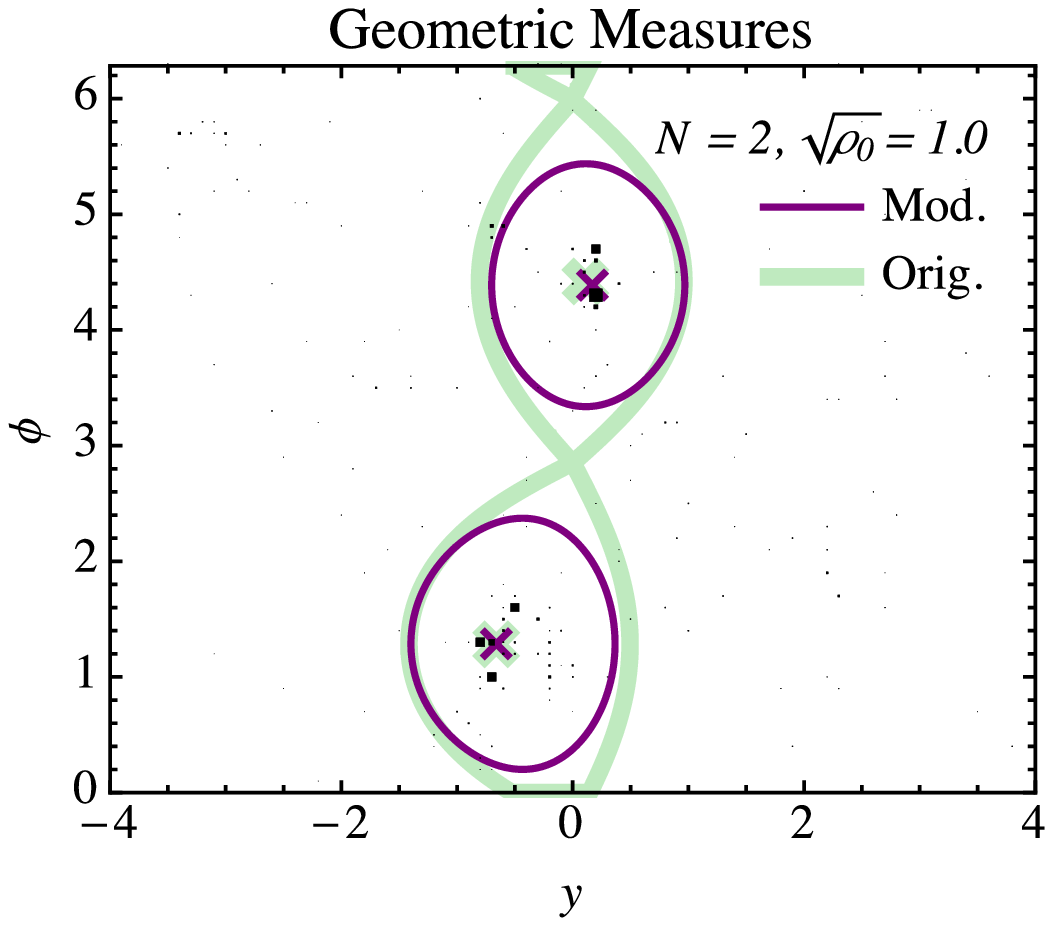}}%
\\
\subfloat[]{\label{fig:jetpicture2:c}%
\includegraphics[width=.49\columnwidth]{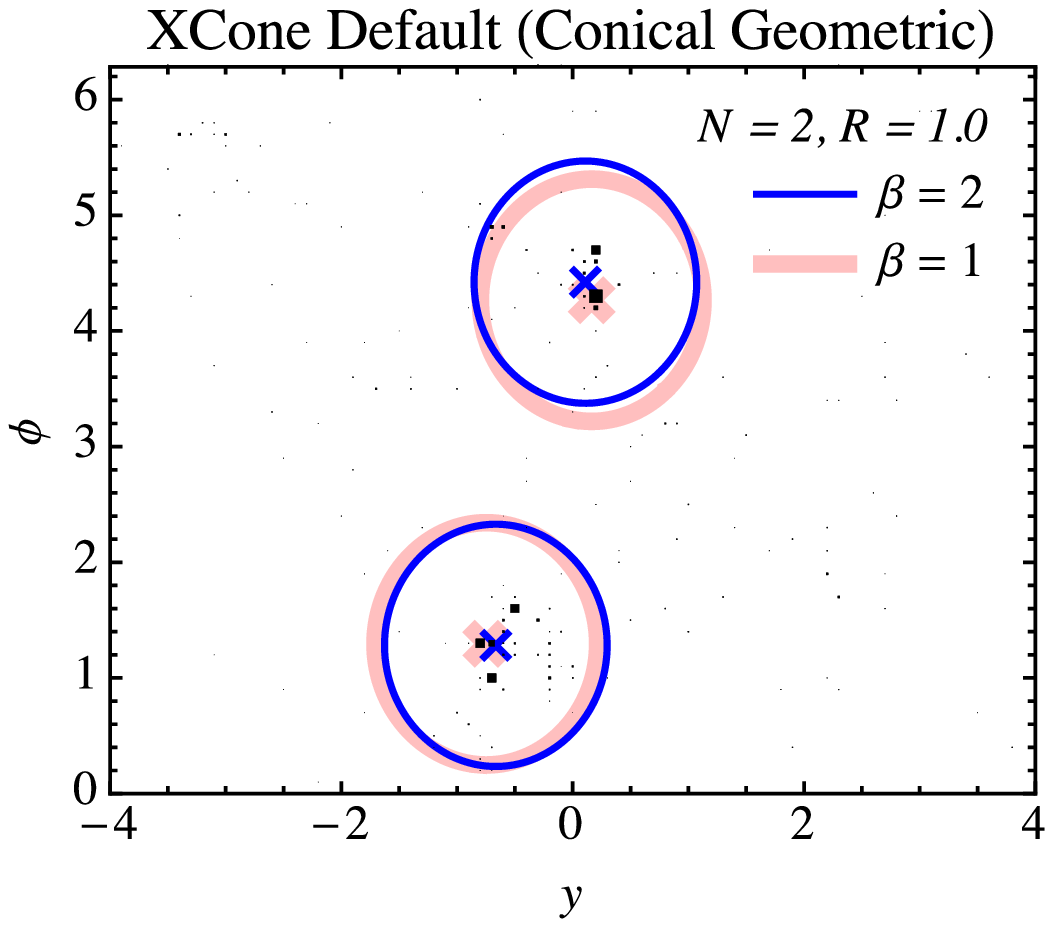}}

\caption{Same $t\bar t$ event as in \Fig{fig:jetpicture}, but for $N = 2$ and $R = 1.0$.  (a) The conical measure yields exactly circular jet regions for widely-separated jets. (b) The geometric measure exhibits cusps at $y = 0$ which are smoothed out with the modified geometric measure.  (c) The XCone default ($\beta = 2$) yields jets centered along the total jet momentum while the recoil-free default ($\beta = 1$) yields jets centered along the hardest cluster within the jet.}
\label{fig:jetpicture2}
\end{figure}

\begin{figure}
\centering
\subfloat[]{\label{fig:jetpicture3:a}%
\includegraphics[width=.49\columnwidth]{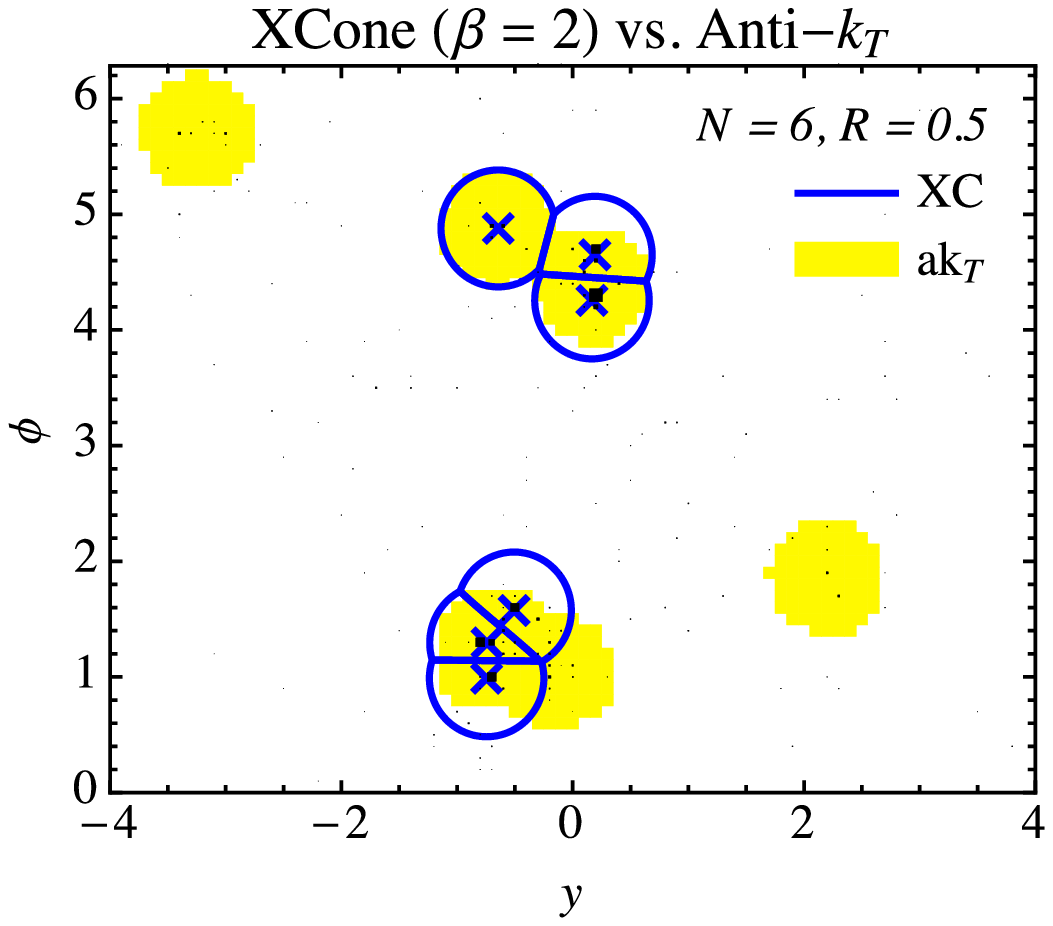}}%
\hfill%
\subfloat[]{\label{fig:jetpicture3:b}
\includegraphics[width=.49\columnwidth]{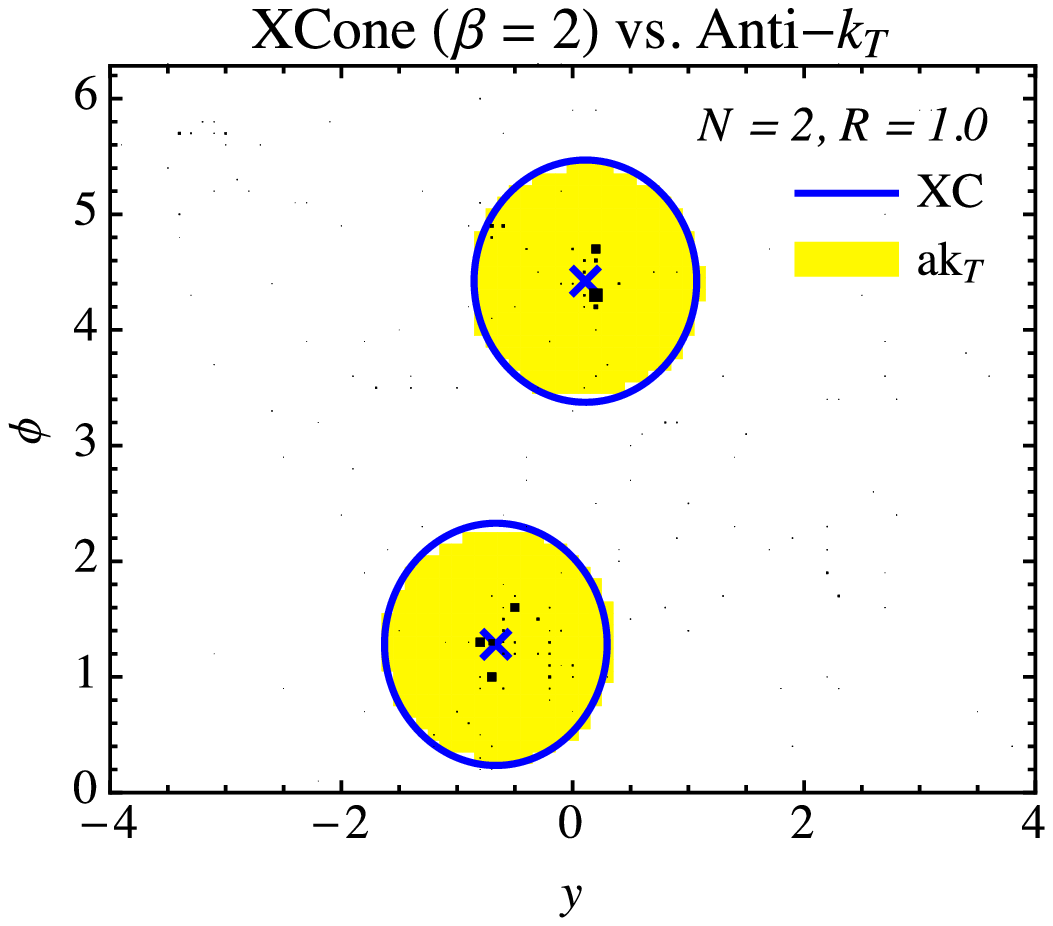}}%
\caption{Comparison between the XCone default ($\beta = 2$) and anti-$k_T$, using the same $t\bar t$ events as \Figs{fig:jetpicture}{fig:jetpicture2}.  (a)  Unlike anti-$k_T$ which merges jet regions closer in angle than $\approx R$, XCone allows such jet regions to remain split.  (b) For widely-separated jets, XCone yields nearly identical jet regions to anti-$k_T$.}
\label{fig:jetpicture3}
\end{figure}

\section{Choice of Measure}
\label{sec:measure}

As already mentioned, every choice of jet and beam measure defines some kind of $N$-jettiness jet algorithm.  We now review previous measures in the literature en route to explaining the logic behind the new XCone default measure.  Example jet regions found from some of these measures are shown in \Figs{fig:jetpicture}{fig:jetpicture2} for a boosted top event from the BOOST 2010 event sample \cite{Abdesselam:2010pt}. In \Fig{fig:jetpicture3}, we show a comparison between the XCone default and the anti-$k_T$ algorithm~\cite{Cacciari:2008gp}.  While XCone and anti-$k_T$ are very similar for widely separated jets as in \Fig{fig:jetpicture3:b}, they behave quite differently when the jets are close together as in \Fig{fig:jetpicture3:a}. 
A more extensive discussion and anti-$k_T$ comparison can be found in the companion paper~\cite{Thaler:2015xaa}.

\subsection{The Conical Measure}

The first conical $N$-jettiness measure was proposed in \Ref{Thaler:2011gf}:\footnote{Strictly speaking, the measure in \Ref{Thaler:2011gf} has an extra rapidity cut parameter.}
\be
\label{eq:defmeasure}
\boxed{\text{Conical Measure}} \qquad\qquad
\begin{aligned}
\rho_{\jet}(p_i,n_A) &= p_{Ti}\, \biggl(\frac{R_{iA}}{R}\biggr)^\beta
\,,\\
\rho_{\beam}(p_i) &= p_{Ti}
\,,\end{aligned}
\ee
where
\begin{equation}
R_{iA} = \sqrt{(y_i - y_A)^2 + (\phi_i - \phi_A)^2}
\end{equation}
is the distance between $p_i$ and $n_A$ in the rapidity-azimuth plane, and $\beta$ is an angular weighting exponent.  The parameter $R$ acts like the jet radius in a cone algorithm, since particle $i$ can only be clustered into jet $A$ if $\rho_{\jet}(p_i,n_A) < \rho_{\beam}(p_i)$, which is equivalent to $R_{iA} < R$.  Thus, the measure in \Eq{eq:defmeasure} yields jets that are exact circles with radius $R$ in the rapidity-azimuth plane, as shown in \Fig{fig:jetpicture2:a}, unless two jet axes are closer than $R$. When two or more axes are closer than $R$ to each other, the jet regions are determined by Voronoi partitioning (i.e.~nearest neighbor).  This yields ``clover jet'' configurations as shown in \Fig{fig:jetpicture:a}.

For small $R$, $\Tau_N$ is dominated by the beam measure, which is just the unclustered $p_T$ in an event.  Thus, this measure typically finds the $N$ jets with the largest $p_T$ in an event.  By adjusting the exponent $\beta$, the jet axis can be varied to point along the jet direction ($\beta = 2$, ``mean'') or along the hardest cluster inside a jet ($\beta = 1$, ``median''), see also \Refs{Thaler:2011gf,Larkoski:2014uqa,Larkoski:2014bia}.

Naively, the conical measure might seem to be the only measure yielding conical jets, since any change to the measure would affect the competition between $\rho_\jet$ and $\rho_\beam$ and change the style of the event partitioning. One can maintain conical jets, however, if one deforms \Eq{eq:defmeasure} via
\be
\label{eq:fmeasure}
\boxed{\text{General Conical Measure}} \qquad\qquad
\begin{aligned}
\rho_{\jet}(p_i,n_A) &= p_{Ti}\, f(p_i)\, \biggl(\frac{R_{iA}}{R}\biggr)^\beta
\,,\\
\rho_{\beam}(p_i) &= p_{Ti}\, f(p_i)
\,,\end{aligned}
\ee
where $f(p_i)$ is any dimensionless function of the particle four-momentum.  This measure still returns exactly conical jets with overlapping jets still having Voronoi partitioning, because the factor of $f(p_i)$ drops out when comparing $\rho_{\jet}$ to $\rho_{\beam}$ or when comparing two different $\rho_{\jet}$. While the partitioning for given axes does not depend on $f(p_i)$, the $f(p_i)$ factor does play a role in determining the overall $\Tau_N$ minimum in \Eq{eq:mincriteria}. So the final jets will have different axes depending on the choice of $f(p_i)$.  We will exploit this possibility when defining the conical geometric measure in \Sec{subsec:dotproduct}.

\subsection{The Geometric Measure}

A variety of $N$-jettiness measures were proposed and studied in \Refs{Jouttenus:2011wh,Jouttenus:2013hs}.  For the purposes of defining a cone jet algorithm, the most promising choice is the geometric measure:
\be
\label{eq:geomeasure}
\boxed{\text{Geometric Measure}} \qquad\qquad
\begin{aligned}
\rho_{\jet}(p_i,n_A) &= \frac{n_A \cdot p_i}{\rho_0}
\,,\\
\rho_{\beam}(p_i) &= \min \{n_a \cdot p_i,\, n_b \cdot p_i\}
\,,\end{aligned}
\ee
where $n_{a,b} = \{1,0,0,\pm 1\}$ and the $z$-direction is the beam direction, such that
\begin{equation}
\min \{n_a \cdot p_i,\, n_b \cdot p_i\} = p_i^0 - \abs{p_i^3} = m_{Ti} e^{-\abs{y_i}}
\,.\end{equation}
Here, $m_{Ti} = \sqrt{p_{Ti}^2 + m_i^2}$, $y_i$ is the rapidity, and this is the form given in \Tab{tab:measures}.

The presence of the $n \cdot p_i$ dot product in the jet and beam measures is very natural from a theoretical perspective, since it makes the measure linear in both $p_i$ and $n$. The linearity in the jet axes $n_A$ implies that the total jet three-momentum is exactly aligned with the axis direction $\vec n_A$ (see \Sec{subsec:update}). The linearity in $p_i$ implies simple factorization properties for $\Tau_N$ and also tends to make perturbative calculations much simpler (see e.g.~\Refs{Jouttenus:2011wh, Bauer:2011uc, Jouttenus:2013hs, Kang:2014qba, Boughezal:2015eha, Gaunt:2015pea}). For this reason all $N$-jettiness calculations so far which involve initial state hadrons have been based on measures linear in $p_i$, like the geometric measure. 

Despite the presence of the dot product $n_A\cdot p_i$, the geometric measure actually behaves quite similarly to the conical measure.\footnote{In the context of recursive clustering algorithms, this dot-product form was also mentioned as an option in \Ref{Catani:1993hr}.}
To see this, note that the momenta $p_i$ and lightlike axes $n_A$ can be expressed as
\begin{align}
p_i &= \bigl\{m_{Ti} \cosh y_i,\, \vec{p}_{Ti},\, m_{Ti} \sinh y_i \bigr\}
\,,&
p_{Ti} &\equiv \abs{\vec{p}_{Ti}}
\,,\\*
n_A &= \bigl\{1,\, \vec{n}_{TA},\, \tanh y_A \bigr\}
\,,&
n_{TA} &\equiv \abs{\vec{n}_{TA}} = \frac{1}{\cosh y_A}
\,,\end{align}
and their dot product is given by
\be \label{eq:dotproduct}
\frac{n_A \cdot p_i}{n_{TA} \, p_{Ti}} = \frac{m_{Ti}}{p_{Ti}} \cosh(y_i - y_A) - \cos(\phi_i - \phi_A)
\,.\ee
In the limit of small angles and for massless particles we thus have
\begin{equation} \label{eq:dotproductapprox}
\rho_{\jet}(p_i,n_A) = \frac{n_A \cdot p_i}{\rho_0} \approx \frac{p_{Ti}}{2\cosh y_A}\, \frac{R_{iA}^2}{\rho_0}
\,.\end{equation}
Hence, the $\rho_{\jet}$ for the geometric measure acts similarly to the general conical measure in \eq{fmeasure} with $\beta = 2$ and $f(p_i) = 1/(2 \cosh y_i)$, at least to the extent that $\cosh y_i \approx \cosh y_A$. This also shows that the parameter $\rho_0$ in the geometric measures controls the size of the jet regions with roughly $\rho_0 \simeq R^2$.

Since the geometric measure does not take the precise form of \Eq{eq:fmeasure}, it yields football-like jets in the central region with cusps at $y = 0$, which get accentuated for larger jets as shown by the green thick lines in \Figs{fig:jetpicture:b}{fig:jetpicture2:b}. For overlapping jets, it produces similar clover jets due to the competition between the $\rho_{\jet}$ for different jets.

Although not as extreme as the jet shapes obtained with an invariant mass measure (see \Ref{Jouttenus:2011wh}), these cusps in the jet boundaries are somewhat unnatural for experimental applications. Since the shape of the jet regions is determined by the competition between $\rho_{\jet}$ and $\rho_{\beam}$, we can modify the geometric measure to yield more conical jets by introducing an explicit compensating factor of $f(p_i) = 1/(2 \cosh y_i)$ in the beam measure:
\be
\label{eq:modgeomeasure}
\boxed{\text{Modified Geometric Measure}} \qquad\qquad
\begin{aligned}
\rho_{\jet}(p_i,n_A) &= \frac{n_A \cdot p_i}{\rho_0}
\,, \\
\rho_{\beam}(p_i) &= \frac{m_{Ti}}{2 \cosh y_i}
\,.\end{aligned}
\ee
With the approximations in \eq{dotproductapprox} and $\cosh y_i \approx \cosh y_A$ this modified measure is now approximately the same as \Eq{eq:fmeasure} with $\beta = 2$ and $f(p_i) = 1/(2 \cosh y_i)$. Hence, it yields reasonably conical jets also in the central region, as shown by the purple lines in \Figs{fig:jetpicture:b}{fig:jetpicture2:b}.  This corresponds to only a slight modification of the geometric beam measure, since close to the beam axes, i.e.\ for large $y_i$, we have
\be
\label{eq:mintoinv}
\frac{m_{Ti}}{2 \cosh y_i} \to m_{Ti}\, e^{-\abs{y_i}}
\,.\ee
This implies that the modified geometric measure has very similar factorization properties as the geometric measure, which we will return to
in \Sec{sec:factorization}.  The use of $1/(2\cosh y_i)$ to replace $e^{-|y_i|}$ is the same as the well-known distinction between using  $C$-parameter~\cite{Parisi:1978eg,Donoghue:1979vi} and thrust~\cite{Farhi:1977sg} event shapes to describe the narrow dijet limit in $e^+e^-$ collisions, see e.g.~\cite{Catani:1998sf,Gardi:2003iv,Korchemsky:2000kp,Hoang:2014wka}.

While we can roughly associate $\rho_0 \simeq R^2$, the jet area itself still differs from $\pi R^2$, especially for larger $R$ and away from central jet rapidities. To enforce jets of a constant jet area, regardless of the jet rapidity and jet boundary, \Ref{Jouttenus:2013hs} also introduced a geometric-$R$ measure where the jet measure is rescaled by a rapidity-dependent factor to maintain $\pi R^2$ jet areas for widely-separated jets:
\be
\label{eq:geoRmeasure}
\boxed{\text{Geometric-$R$ Measure}} \qquad\qquad
\begin{aligned}
\rho_{\jet}(p_i,n_A) &= \frac{1}{\rho(R,y_A)} \, n_A \cdot p_i
\,,\\
\rho_{\beam}(p_i) &= m_{Ti} e^{-\abs{y_i}} 
\,.\end{aligned}
\ee
Here, $\rho(R,y_A)$ is given in terms of the the integral $I_0(\alpha,\beta)$ from~\cite{Jouttenus:2011wh}  which determines the geometric jet area (for nonoverlapping jets) via the transcendental equation~\cite{Stewart:2014nna}
\begin{align}
  I_0\Big(\frac{a_+}{2\rho},\frac{a_-}{2\rho}\Big) + 
    I_0\Big(\frac{a_-}{2\rho},\frac{a_+}{2\rho}\Big) = R^2 
  \,, \qquad\quad
  a_\pm = 1 \pm \tanh y_A \,.
\end{align}
Numerical results for $\rho$ were given in \Ref{Jouttenus:2013hs}. The same modifications as above lead to the modified geometric-$R$ measure
\be
\label{eq:modgeoRmeasure}
\boxed{\text{Modified Geometric-$R$ Measure}} \qquad\qquad
\begin{aligned}
\rho_{\jet}(p_i,n_A) &= \frac{1}{\rho_C(R,y_A)}\, n_A \cdot p_i
\,,\\
\rho_{\beam}(p_i) &= \frac{m_{Ti}}{2 \cosh y_i}
\,,\end{aligned}
\ee
where $\rho_C(R, y_A)$ is different than in \eq{geoRmeasure} due to the difference in the beam measures.

In all of the above cases, the rapidity suppression in the beam measures at large rapidities makes $\Tau_N$ much less sensitive to the forward region. This means the $\Tau_N$ minimization effectively corresponds to minimizing a rapidity-weighted sum of unclustered $p_T$ (i.e.\ the unclustered beam thrust~\cite{Stewart:2009yx} or ``beam $C$-parameter'' contribution). As a result, the algorithm will dominantly identify central jets over forward jets, which could have interesting applications, e.g.\ when one wants to avoid picking up forward jets from initial-state radiation. Corresponding forward-insensitive rapidity-weighted jet vetoes have been discussed recently in \Ref{Gangal:2014qda}.

\subsection{The Conical Geometric Measure}
\label{subsec:dotproduct}

Combining the lessons of the conical and geometric measures, we now introduce the conical geometric measure which aims to combine their advantages.  For a specific choice of parameters, this will be the XCone default measure.

Like the conical measure, we want a measure that returns (nearly) conical jets, and we also want a parameter $\beta$ in the jet measure to adjust the behavior of the jet axes. Like the geometric measure, we want a measure that depends on the dot products between lightlike axes and particles, since that is the simplest distance to use in theoretical calculations, and can be made linear in the particle momentum (here by choosing $\beta=2$). These requirements lead us to
\be
\label{eq:dotproductmeasure}
\boxed{\text{Conical Geometric Measure}} \qquad\quad
\begin{aligned}
\rho_{\jet}(p_i,n_A) &= \frac{p_{Ti}}{(2 \cosh y_i)^{\gamma - 1}} \biggl(\frac{2 n_A \cdot p_i}{n_{TA} \, p_{Ti}}\,\frac{1}{R^2} \biggr)^{\beta/2}
,\\
\rho_{\beam}(p_i) &= \frac{p_{T i}}{(2 \cosh y_i)^{\gamma - 1}}
\,,\end{aligned}
\ee
where again $n_{TA} = 1/ \cosh y_A$.  In the jet measure, we recognize the last factor in parentheses as the approximate form for $R_{iA}$ in \Eq{eq:dotproduct}, which now yields jets that are very nearly conical. The $\beta$ factor acts just like the $\beta$ factor in the conical measure.  For additional flexibility, we have chosen a common $f(p_i) = (2 \cosh y_i)^{1 - \gamma}$ in the beam and jet measures. This multiplicative factor affects the axes found by minimization, but not the beam and jet regions. It is parametrized by $\gamma$, such that for $\gamma = 1$ this reproduces the beam measure of the conical measure while for $\gamma = 2$ this is closely analogous to the beam measure of the modified geometric measures.

\begin{figure}
\centering
\includegraphics[width=.45\columnwidth]{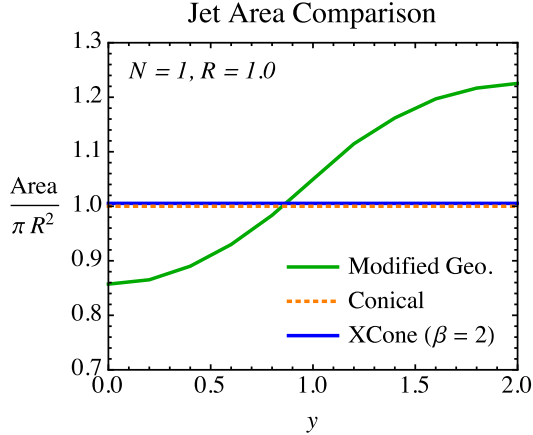}
\caption{Comparison of the analytic jet areas for a single jet ($N = 1$).  Unlike the modified geometric measure, the conical geometric measure (here shown for the XCone default of $\beta = 2$) has uniform jet areas as a function of rapidity.  For $R \lesssim 1.0$, this area is within 1\% of $\pi R^2$ from the conical measure.}
\label{fig:jetarea}
\end{figure}

There is additional freedom in defining the conical geometric measure that we will not exploit in this paper.  For example, we could multiply the jet or beam measures by any function of
\be
\label{eq:mpfactorambiguity}
\frac{m_{Ti}}{p_{Ti}}
\,,\ee
which would give slightly different behavior for massive hadrons.  In the jet measure, we could multiply by any function of
\be
\label{eq:yfactorambiguity}
\frac{\cosh y_i}{\cosh y_A}
\,,
\ee
since this quantity is nearly one for narrow jets. For example, the modified geometric measure is reproduced exactly by taking $\beta = \gamma = 2$ and in addition multiplying the beam and jet measures by $m_{Ti}/p_{Ti}$ and $\cosh y_i / \cosh y_A$, respectively.  These choices are somewhat analogous to the choice of recombination schemes in recursive jet algorithms, since they are irrelevant for infinitely narrow cones and massless inputs.
That said, for $\beta=2$, $\gamma=2$ the factor of $\cosh y_A/ \cosh y_i$ that appears in the conical geometric measure relative to the (modified) geometric measure ensures that the jet area is very close to $\pi R^2$ even for relatively forward jets, as shown in \Fig{fig:jetarea}.

\subsection{The XCone Default Measure}

For LHC applications, our recommended XCone default is the conical geometric measure with $\beta = 2$ and $\gamma = 1$:
\be
\label{eq:xconemeasure}
\boxed{\text{XCone Default Measure  ($\beta = 2$)}} \qquad\qquad
\begin{aligned}
\rho_{\jet}(p_i,n_A) &= \frac{2 \cosh y_A}{R^2} \, n_A \cdot p_i
\,,\\
\rho_{\beam}(p_i) &= p_{T i}
\,.\end{aligned}
\ee
By choosing $\gamma = 1$, the beam measure is the same as the conical measure, so minimizing $\Tau_N$ minimizes the unclustered $p_T$.  By choosing $\beta = 2$, the jet axis (approximately) aligns with the total three-momentum of the jet, as is typical for traditional stable cone algorithms.  Note that the jet measure is linear in $p_i$, as desired for theoretical calculations.  In  \Fig{fig:jetarea} we show that the active area of XCone jets is very nearly $\pi R^2$ for well-separated jets, see also \Ref{Thaler:2015xaa}.

Alternatively, in cases where recoil-sensitivity  \cite{Catani:1992jc, Dokshitzer:1998kz, Banfi:2004yd, Larkoski:2013eya} is an issue (such as in high pileup environments \cite{Larkoski:2014bia}) we can use $\beta = 1$ and $\gamma = 1$:
\be
\label{eq:recoilfreexconemeasure}
\boxed{\text{Recoil-Free Default Measure  ($\beta = 1$)}} \qquad\qquad
\begin{aligned}
\rho_{\jet}(p_i,n_A) &= \sqrt{\frac{2 \cosh y_A}{R^2} \, p_{T i} \, n_A \cdot p_i }
\,,\\
\rho_{\beam}(p_i) &= p_{T i}
\,.\end{aligned}
\ee
Here, the jet center aligns approximately along the broadening axis of the jet \cite{Thaler:2011gf,Larkoski:2014uqa}, which is the axis that minimizes the summed transverse momentum relative to it.  This is similar to finding the ``median'' jet energy and the jet axis tends to point along the most energetic cluster within a given jet.  Again, the jet area is approximately $\pi R^2$.

These XCone default measures are the basis for our LHC case studies in the companion paper \cite{Thaler:2015xaa}, where we find that both $\beta = 2$ and $\beta = 1$ give comparable results for jet reconstruction (in the absence of jet contamination). The jet regions for XCone default are shown in \Figs{fig:jetpicture:c}{fig:jetpicture2:c}. With a single energetic cluster inside a jet, the difference between $\beta = 2$ and $\beta = 1$ is very small (again in the absence of jet contamination), analogous to the way that the mean and median of a peaked distribution are very similar.  This is shown in \Fig{fig:jetpicture:c}. When a jet has substructure, the ``mean'' ($\beta = 2$) and ``median'' ($\beta = 1$) axes are offset, as shown in \Fig{fig:jetpicture2:c} for the same event with $N = 2$. One can also see that for larger jet radius, the jet regions are slightly elongated along the azimuthal direction compared to the rapidity direction. This arises because of the trigonometric functions in \eq{dotproduct}. In \Ref{Thaler:2015xaa} it is mentioned that this deformation from exact circles yields slightly improved performance when reconstructing invariant-mass peaks.

\section{Details of the XCone Algorithm}
\label{sec:details}

For a given $N$-jettiness measure entering in \eq{tauNdef}, we need to implement the minimization procedure in \eq{mincriteria} to determine the jet axes $n_A$.  In general, the only guaranteed method to find the global minimum of $\Tau_N$ is to test by brute force all possible partitions of the final-state particles into $N$ jet regions and one beam region.  Since this is computationally prohibitive, our aim is to find good approximations of the global minimum by relying on methods that strictly speaking only find local minima of $\Tau_N$. Even if the algorithm does not find a guaranteed global $\Tau_N$ minimum, as long as all steps are fully specified and IRC safe, it still represents a well-defined exclusive cone algorithm which retains the key features of the $N$-jettiness partitioning according to the specified jet and beam measures.

Throughout this section, we restrict ourselves to the case $\gamma = 1$, which is currently implemented in the XCone code and is also used by the default measures.

\subsection{One-Pass Minimization}
\label{subsec:onepass}

For the conical measure in \Eq{eq:defmeasure}, \Ref{Thaler:2011gf} introduced a modification of Lloyd's method \cite{Lloyd82leastsquares} that finds a local minimum of $\Tau_N$ for $1 < \beta < 3$.  We can adopt a similar strategy for more general measures.

Our minimization algorithm proceeds as follows, with more details given below:
\begin{enumerate}
  \item[1)] \emph{Find seed axes:} Determine a set of suitable IRC-safe initial axes $n_A$.
  \item[2)] \emph{Assignment:} For fixed axes $n_A$, assign particles to jet and beam regions via $\Tau_N$ partitioning.
  \item[3)] \emph{Update axes:} For fixed partitioning, update axes $n_A$ via $\Tau_N$ minimization.
  \item[4)] If axes have converged then stop, otherwise go back to step 2).
\end{enumerate}
To be IRC safe, this procedure must be fully deterministic. We therefore always perform a one-pass minimization, i.e., the above algorithm is repeated precisely once per event without any stochastic elements (such as random variations in the seed axes). The procedure to determine the seed axes in step 1) is deterministic and IRC safe, as described in \Sec{subsec:minfind}. The seed axes are then iteratively improved to a local minimum of $\Tau_N$ in steps 2) and 3).

In the assignment step 2), the final-state particles are assigned to one of the $N$ jet regions or to the beam region via the $\Tau_N$ partitioning in \Eq{eq:tauNdef} for the current set of fixed trial axes $n_A$.  This step can be easily implemented for any choice of measure as it only depends on the competition between the jet measures $\rho_{\jet}(n_A, p_i)$ for fixed $n_A$ and the beam measure $\rho_{\beam}(p_i)$, so we do not need to discuss it further.

In the update step 3), the axes $n_A$ are improved to minimize the contribution to the $\Tau_N$ value within each jet region, keeping the jet constituents determined by the partitioning in the previous assignment step fixed. Different update steps are needed for different measures, since there is no general procedure to find the axes $n_A$ that minimize $\sum_i \rho_{\jet}(n_A, p_i)$.\footnote{Even if one does find such a procedure, one has to check on a case-by-case basis whether the assignment/update iteration actually converges when using it. Some pathological cases were discussed in \Ref{Thaler:2011gf}.}  Once an appropriate update step is found, the assignment and update steps can be iterated until the axes converge to within some specified accuracy.  In \Sec{subsec:update}, we describe a general update step that works well for the measures studied in this paper.

As discussed in \Ref{Thaler:2011gf}, these one-pass minimization procedures are quite effective for $N$-subjettiness, often converging to the global minimum.  There are additional complications, however, for $N$-jettiness.  The reason is that $N$-jettiness has a beam region, and particles in the bulk of the beam region are insensitive to small changes to the location of the jet axes $n_A$.  Even minimization routines that try to go ``uphill''  to escape local minima may never find the optimal jet axes. Given that $\Tau_N$ corresponds roughly to the unclustered $p_T$ in an event (for $\gamma = 1$), failing to find a decent $\Tau_N$ minimum means that one will identify too many soft jets.  Therefore, for XCone to be a practical jet algorithm, one has to find a good set of seed axes for one-pass minimization.  In \Sec{subsec:minfind}, we show how to find  such seed axes by utilizing recursive clustering algorithms.

Another possibility to further improve the $\Tau_N$ minimization is by running the above (or any other) exclusive jet algorithm to find $N+n$ jets. Starting from these, one can then perform the remaining partitioning into $N$ jets by explicitly testing all possible combinatorial options to find the best minimum.  This option is available in the XCone code, though not recommended by default for reasons of speed. One advantage of this strategy is that it reduces to the exact $\Tau_N$ minimization for up to $N+n$ final-state particles. This makes it convenient for fixed-order calculations up to N$^n$LO, where the one-pass minimization with seed axes could induce rather complicated boundaries in the phase-space integrations.

\subsection{Update Step for General Measures}
\label{subsec:update}

We now construct a general update step that converges to a local minimum of $\rho_{\jet}(n_A, p_i)$ for a fixed set of jet constituents with momenta $p_i$.  This approach works for a wide variety of jet measures, including the XCone defaults.

To motivate our general procedure, we start with the special case of the (modified) geometric measure, where finding a local minimum of $\rho_{\jet}$ is particularly straightforward.  Within a given jet region $A$, we want to find the axis $n_A$ that minimizes
\be
\label{eq:nicemin}
\sum_{i \in A} n_A \cdot p_i = n_A \cdot \Bigl(\sum_{i \in A} p_i \Bigr) \equiv n_A \cdot p_A
\,,\ee
where $p_A = \sum_{i \in A} p_i$ is the total four-momentum of all jet constituents.  Introducing a Lagrange multiplier $\lambda$ (as in \cite{Thaler:2015uja}), the quantity
\be
n_A \cdot p_A + \lambda (\vec{n}_A^2 - 1)
\ee
is minimized for
\begin{align}
n_A = \Bigl\{1, \frac{\vec{p}_A}{|\vec{p}_A|} \Bigr\}
\qquad\text{with}\qquad
\vec p_A = \sum_{i \in A} \vec p_i
\,,\end{align}
such that the jet axis $\vec n_A$ exactly aligns with the total three-momentum of the jet.  Thus, minimizing the modified geometric measure is equivalent to finding $N$ mutually stable (Voronoi-bounded) cones.  In the same way, any measure of the form
\be
\label{eq:rhojetniceform}
\rho_{\jet}(n_A, p_i) = n_A \cdot p_i \, f(p_i)
\ee
will be minimized by
\be
n_A = \Bigl\{1, \frac{\vec{q}_A}{|\vec{q}_A|} \Bigr\}
\qquad\text{with}\qquad
\vec q_A = \sum_{i \in A} \vec p_i \, f(p_i),
\ee
where $\vec q_A$ is the effective total three-vector of the $f$-weighted jet constituents. For these cases, one-pass minimization will terminate in a finite number of assignment/update steps.

The conical geometric measure does not take the form of \Eq{eq:rhojetniceform}, but rather takes the more general form
\be
\rho_{\jet}(n_A, p_i) = n_A \cdot p_i \, g(p_i, n_A),
\ee
where the jet measure has nonlinear dependence on $n_A$.  This means that the jet axis and the jet three-momentum do not in general align.  For the XCone default measure in particular, the extra factor of $\cosh y_A$ in the jet measure means that there is an offset between the axis and the momentum proportional to the jet mass.  Thus, we cannot directly use the above stable-cone finding logic to minimize $\rho_{\jet}$.  Instead, as in \Ref{Thaler:2011gf}, we can define an update step based on the previous $n_A$ value:%
\footnote{For practical purposes, it is sometimes necessary to include an ``effective mass'' term by changing $n_A \cdot p_i \to n_A \cdot p_i + \epsilon $ with small $\epsilon$ to avoid potential divide-by-zero errors.}
\be
n_A^{\text{new}} =\Bigl\{1, \frac{\vec{q}_A}{|\vec{q}_A|} \Bigr\}
 \qquad\text{with}\qquad
 \vec q_A = \sum_{i \in A} \vec p_i \, g(p_i, n^{\rm old}_A).
\ee
As long as the dependence on $n_A$ is mild enough (roughly $1 \le \beta < 3$ for the conical geometric measure), this procedure will converge within a desired accuracy in a reasonable number of assignment/update iterations, and we adopt this strategy for the XCone default measures. (In practice, due to the presence of local minima, the one-pass minimization may converge to a higher value of $\Tau_N$ than the original seed axes value.  For this reason, we always return the smallest $\Tau_N$ value and associated axes seen among all update steps.)

\subsection{Seed Axes for One-Pass Minimization}
\label{subsec:minfind}

Recursive clustering algorithms are particularly effective to find seed axes for one-pass minimization.  When run in exclusive mode, a recursive clustering algorithm  returns exactly $N$ jets which can then be interpreted as $N$ lightlike seed axes.  In fact, the axes are often so good in practice that the iterative improvement step is unnecessary.  One could even imagine a more general strategy that separates jet axis finding (here using recursive clustering) from jet region finding (here using $N$-jettiness partitions), and we plan to pursue this possibility in future work.  Unlike generic cluster optimization, recursive clustering algorithms are computationally efficient, and this efficiency is inherited by our XCone implementation (at the expense of only guaranteeing a local $\Tau_N$ minimum).

For the conical geometric measures with $\gamma = 1$, including the XCone defaults, good seed axes can be found by running the generalized $k_{T}$ clustering algorithm with a generalized $E_{t}$ recombination scheme.  The generalized $k_{T}$ clustering measure \cite{Cacciari:2008gp,Cacciari:2011ma} is parametrized by an exponent $p$ and a jet radius $R$:
\be
d_{ij} = \mathrm{min}\left(p_{Ti}^{2p}, p_{Tj}^{2p}\right)\frac{R_{ij}^2}{R^2}, \qquad  d_{iB} = p_{Ti}^{2p},
\label{eq:genkt}
\ee
where $p = 1$ is the $k_T$ algorithm \cite{Catani:1993hr,Ellis:1993tq} and $p = 0$ is the Cambridge/Aachen algorithm \cite{Dokshitzer:1997in,Wobisch:1998wt,Wobisch:2000dk}.  The generalized $E_{t}$ recombination scheme is parametrized by an energy-weighting power $\delta$, such that one obtains a massless recombined four-momentum $p_r$ given by
\be
p_{Tr} = p_{Ti} + p_{Tj}, \qquad
\phi_{r} = \frac{p_{Ti}^{\delta}\phi_i + p_{Tj}^{\delta}\phi_j}{p_{Ti}^{\delta} + p_{Tj}^{\delta}},\qquad
\eta_{r} = \frac{p_{Ti}^{\delta}\eta_i + p_{Tj}^{\delta}\eta_j}{p_{Ti}^{\delta} + p_{Tj}^{\delta}},
\label{eq:genrecomb}
\ee
where $\delta = 1$ is the original $E_t$ scheme, $\delta = 2$ is the $E_t^2$ scheme \cite{Catani:1993hr,Butterworth:2002xg}, and $\delta = \infty$ is the winner-take-all scheme \cite{Bertolini:2013iqa,Larkoski:2014uqa,Salambroadening}.

\begin{figure}[t]
\centering
\subfloat[]{\label{fig:heuristic:a}%
\includegraphics[width = .49\columnwidth]{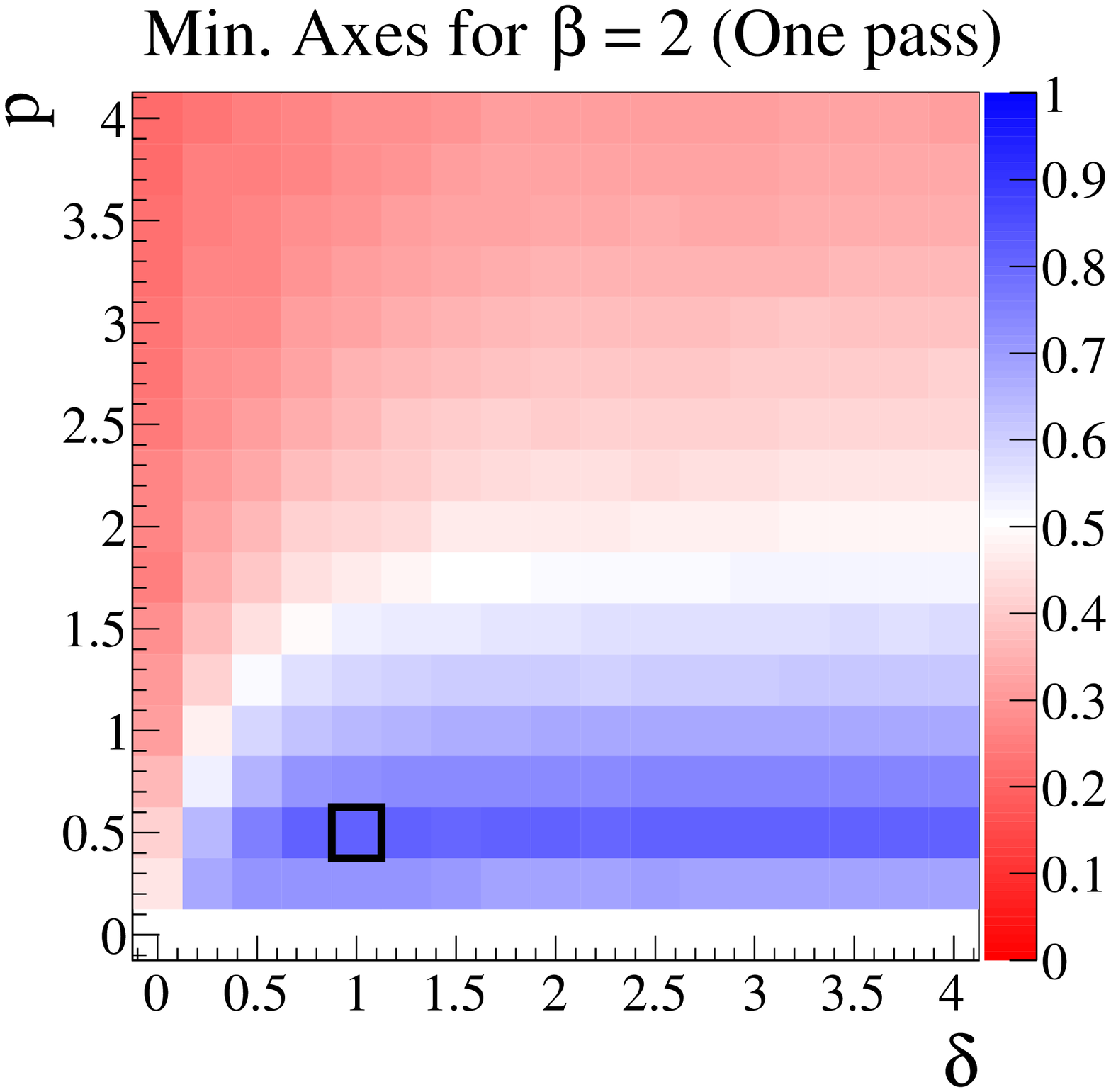}}%
\hfill%
\subfloat[]{\label{fig:heuristic:b}%
\includegraphics[width = .49\columnwidth]{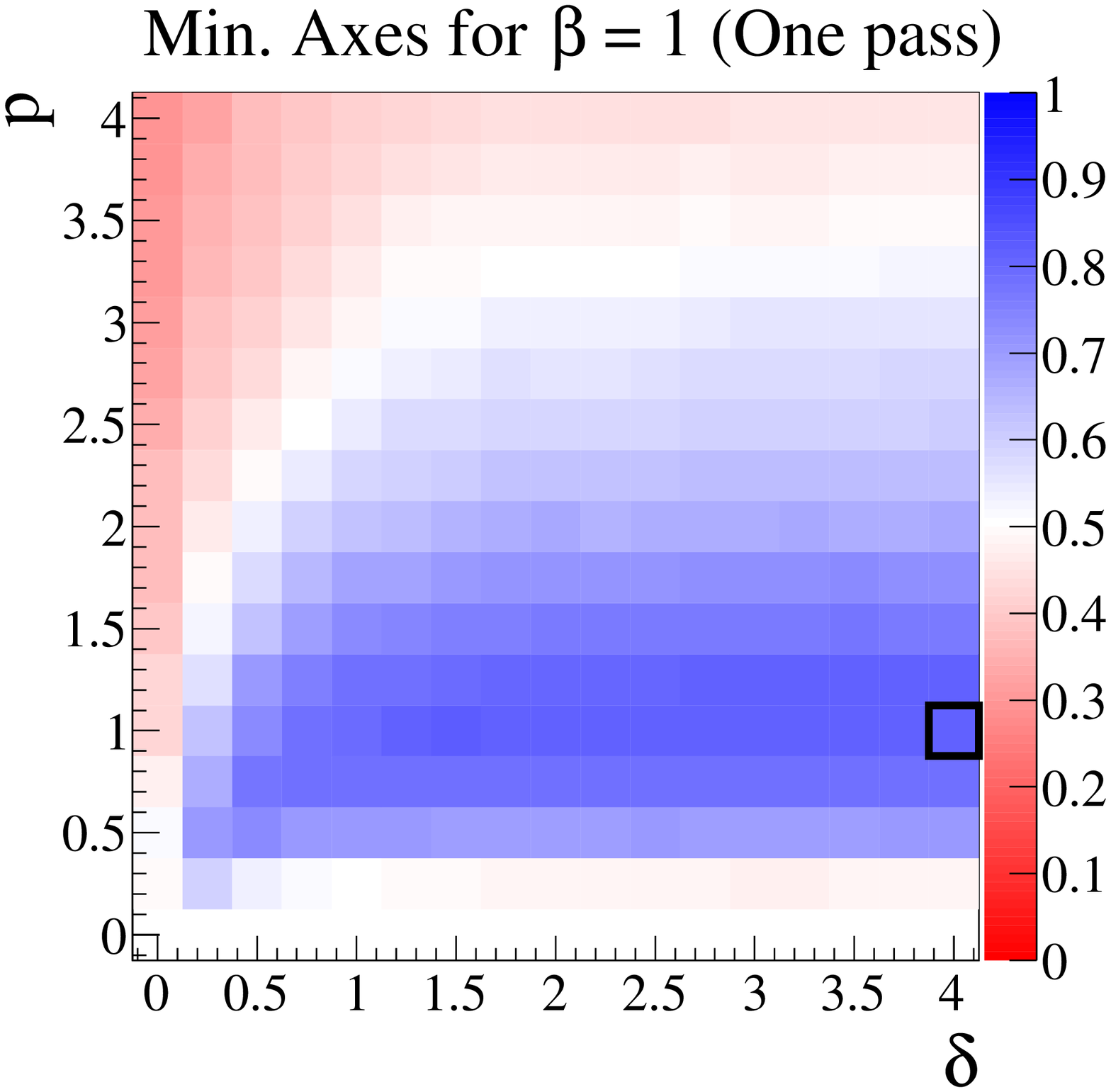}}%
\caption{Fraction of events where all XCone jets from one-pass minimization starting from generalized $k_T$ jet axes as seeds align with the axes from global $\Tau_N$ minimization.  This is for the BOOST 2010 top sample (Herwig 6.5, $p_{T} \in [500, 600]~\GeV$) \cite{Abdesselam:2010pt}, using the conical geometric measure with $N = 6$ and $R = 0.5$.  (a)  The XCone default ($\beta = 2$).  (b) The recoil-free default ($\beta = 1$).  Here, $p$ and $\delta$ parametrize the generalized $k_T$ metric and recombination scheme, respectively.  The black boxes indicate the preferred values of $p$ and $\delta$ from the heuristic choice in \Eq{eq:pdeltaheuristic} (with $\delta = 4$ indicating $\delta \to \infty$).}
\label{fig:heuristic}
\end{figure}

For finding seed axes, the recommended parameters for $0 < \beta < 2$ are
\be
\label{eq:pdeltaheuristic}
p \simeq \frac{1}{\beta}, \qquad \delta \simeq  \frac{1}{\beta - 1}
\,,\ee
with matching radius parameter $R$.  To understand this heuristic choice, consider starting with a final state of $N + 1$ particles and running one iteration of exclusive generalized $k_{T}$ to find $N$ axes.  For this procedure to give good seed axes for $\Tau_N$ minimization, we want to choose the values of $p$ and $\delta$ that match the behavior of the $N$-jettiness metric as closely as possible.  Essentially, we want $d_{iB}$ to match the beam measure $\rho_{\beam}$, $d_{ij}$ to match the jet measure $\rho_{\jet}$, and the recombination scheme to appropriately place the merged axis in the desired location.

We perform this heuristic analysis for the conical measure, which is a bit easier to understand than the conical geometric measure, though the same conclusions hold.  To match the conical beam measure, generalized $k_{T}$ with $p \ge 0$ already gives the right behavior, since the softest particle farther than $R$ from any other particle is merged with the beam.%
\footnote{In principle, it is possible to also handle the $\gamma \not= 1$ case by further modifications of \Eq{eq:genkt} (such as those proposed in \Ref{Boronat:2014hva}), but we have not attempted that for the present XCone implementation.}

To match the conical jet measure, we want $d_{ij}$ to depend on the combination $p_{Ti} R_{ij}^\beta$, which is achieved for
\be
p = \frac{1}{\beta}
\,.\ee
To match the conical axis behavior, we have to know which axis minimizes the $\Tau_N$ value for a jet region consisting of two particles.  Labeling the two particles 1 and 2 and simplifying to one dimension $\phi$ without loss of generality, we have
\be
\Tau_N \sim p_{T1} |\phi_1 - \phi_A|^\beta + p_{T2} |\phi_2 - \phi_A|^\beta
\,,
\ee
where $\phi_A$ is the location of the axis.  Solving $\df \Tau_N / \df \phi_A = 0$ to find the location of the minimum, we find
\be
\phi_A = \frac{p_{T1}^\delta \phi_1 +  p_{T2}^\delta \phi_2}{p_{T1}^\delta +  p_{T2}^\delta}, \qquad \delta = \frac{1}{\beta -1}
\,,
\ee
which is exactly the generalized $E_t$ recombination scheme.  This is the logic behind the heuristic choice in \Eq{eq:pdeltaheuristic}.

\begin{table*}[t]
\subfloat[]{%
\small%
\begin{tabular}{ccccc}
\hline \hline
$(\beta = 2)$ & Seed axes & One-pass min \\
\hline\hline
Jets & 0.95 & 0.96 \\
\hline
Events ($\ge 4$) & 0.99 & 0.99 \\
\hline
Events ($\ge 5$) & 0.92 & 0.93 \\
\hline
Events (6) & 0.78 & 0.81 \\
\hline \hline
\end{tabular}%
\label{tab:percent_min_beta2}}%
\hfill%
\subfloat[]{%
\small%
\begin{tabular}{ccccc}
\hline \hline
$(\beta = 1)$ & Seed axes & One-pass min \\
\hline\hline
Jets & 0.95 & 0.97 \\
\hline
Events ($\ge 4$) & 0.99 & 0.99 \\
\hline
Events ($\ge 5$) & 0.97 & 0.98 \\
\hline
Events (6) & 0.72 & 0.81 \\
\hline \hline
\end{tabular}%
\label{tab:percent_min_beta1}}%
\caption{Fraction of XCone jets that are aligned with the ``true'' minimum from global $\Tau_N$ minimization using only the seed axes from generalized $k_T$ jets and after one-pass minimization for (a) $\beta = 2$ and (b) $\beta = 1$.  Also shown are the fraction of all events with 4 or more, 5 or more, and all 6 jets aligned with the global minimum.}
\end{table*}

To explicitly validate the choice in \eq{pdeltaheuristic}, we consider a sample of boosted top quarks from the BOOST 2010 report \cite{Abdesselam:2010pt}, using $N = 6$ and $R = 0.5$.  A key feature of this boosted top sample is the presence of initial-state radiation, which generates an additional seventh hard jet in the event, providing a nontrivial test scenario. We first determine by brute force the global $\Tau_N$ minimum, as best as we can, by performing one-pass minimization on a wide range of seed axes.  Next, we perform the one-pass minimization with the generalized $k_T$ jets as seed axes for a range of $p$ and $\delta$ values. For each $p$ and $\delta$, we then count the fraction of events that have all $N = 6$ XCone jet axes within $\Delta R < 0.1$ of the axes found from global $\Tau_N$ minimization.  The results are shown in \Fig{fig:heuristic}, which shows that the choice in \Eq{eq:pdeltaheuristic}, shown by the black boxes, does give the best performance. We also observe that a wide range of $\delta$ values give similar results, while the choice of $p$ is more relevant, especially for $\beta = 2$.

The fraction of aligned XCone jets, as well as the fraction of events where $\geq 4$, $\geq 5$, and all $6$ XCone jets are aligned with the global minimum, both before and after one-pass minimization, are shown in \Tabs{tab:percent_min_beta2}{tab:percent_min_beta1}. Even without one-pass minimization, i.e.\ using the seed axes only, $95\%$ of the individual jets are closely aligned with the global $\Tau_N$ minimization for both $\beta = 2$ and $\beta = 1$.  This suggests that finding local $\Tau_N$ minima from generalized $k_T$ seed axes is a robust procedure that often results in a global $\Tau_N$ minimum.

The presence of additional hard jets from initial-state radiation can of course confuse $N = 6$ jet finding, leading to a roughly 70-80\% success rate for correctly identifying all 6 jets originating from the top decays. It is also not obvious that $\Tau_N$ minimization will necessarily always yield the best boosted top reconstruction, and it might well be that ``failed'' $\Tau_N$ minima are still useful for physics analyses. For a detailed study of the phenomenological aspects we refer to \Ref{Thaler:2015xaa}, which also explores an $N = 2 \times 3$ strategy for this final state.

\section{\boldmath $N$-jettiness Factorization with Various Measures}
\label{sec:factorization}

A key attribute that originally motivated the use of $N$-jettiness is its factorization properties in the limit $\widetilde \Tau_N\to 0$~\cite{Stewart:2010tn}, which greatly simplifies calculations of the corresponding exclusive jet cross sections.  The original $N$-jettiness factorization theorem was derived for active-parton cross sections\footnote{We only consider factorization for active-parton cross sections, initiated by incoming quarks or gluons, in order to avoid the complications associated with the spectator partons present for incoming hadrons, such as Glauber effects~\cite{Collins:1988ig,Gaunt:2014ska,Zeng:2015iba}. When using these active-parton factorization theorems, it is nevertheless often assumed that the initial-state quarks and gluons are determined by standard parton distributions.  For the $N$-jettiness observables, Glauber effects have not been fully treated in the literature.} using techniques from  Soft-Collinear Effective Theory (SCET)~\cite{Bauer:2000ew, Bauer:2000yr, Bauer:2001ct, Bauer:2001yt, Bauer:2002nz}, which we also make use of here. So far, these properties have only been fully studied for situations where the measure is linear in a component of the particle momenta~\cite{Stewart:2010tn,Jouttenus:2011wh,Stewart:2009yx,Kang:2013nha}, which simplifies the objects appearing in the factorization theorem.  The examples studied thus far include the geometric and geometric-$R$ measures in \Tab{tab:measures}.

In this section, we derive the factorization properties for more general measures. We will start with a generic analysis and eventually focus on $\beta = 2$ jet measures. We investigate the impact of the choice of jet axes and different beam measures. We also explain how transverse momentum conservation restricts the range of jet observables that can be calculated using the simplest version of the $N$-jettiness factorization theorem.

\subsection{Separating into Jet and Beam Regions}

Due to the linear sum over particles $i$ in \eq{tauNdef}, $N$-jettiness can be obtained by adding up distinct contributions from the beam and $N$ jet regions $r$
\begin{align} \label{eq:linearregions}
\widetilde \Tau_N 
  &=\sum_r \widetilde \Tau_N^r 
   = \widetilde \Tau_N^{a} + \widetilde \Tau_N^{b} +
   \widetilde \Tau_N^{1} + \dotsb + \widetilde \Tau_N^{N} \,.
\end{align}
If only a single measurement is made on the beams as in \eq{tauNdef}, we can simply use $\widetilde \Tau_N^{a} + \widetilde \Tau_N^{b}=\widetilde \Tau_N^{\rm beam}$ here.  Thus the $N$-jettiness cross section is obtained from the more fundamental cross section which is fully differential in the $\widetilde \Tau_N^i$ for each region,
\begin{align}
 \frac{\df\sigma(X_N)}{\df  \widetilde \Tau_N}  
   &= \int\! \biggl[ \prod_{r} \df\widetilde\Tau_N^{\,r}\biggr]
   \delta\Bigl(\widetilde\Tau_N - \sum_{r} \widetilde\Tau_N^{\,r}\Bigr)
   \frac{\df\sigma(X_N)}{\df\widetilde \Tau_N^{a} \df\widetilde \Tau_N^{b} 
    \df\widetilde\Tau_N^1\dotsb \df\widetilde\Tau_N^N}
  \,,
\end{align}
where the products and sum run over $r=a,b,1,\ldots,N$. Here $X_N$ denotes a set of measurements made on the $N$ signal jets and on other final-state particles like electroweak bosons or nonhadronic decay products which we write as follows
\begin{align} \label{eq:dsigmasingfull}
\frac{\df\sigma(X_N)}{\df\widetilde \Tau_N^{a} \df\widetilde \Tau_N^{b} 
    \df\widetilde\Tau_N^1\dotsb \df\widetilde\Tau_N^N}
&= \int\!\df\Phi_N\, \sum_{\kappa}  s_\kappa\, \frac{\df\sigma_\kappa (\Phi_N)}
   {\df\widetilde \Tau_N^{a} \df\widetilde \Tau_N^{b} 
    \df\widetilde\Tau_N^1\dotsb \df\widetilde\Tau_N^N} \,X_N(\Phi_N)
\,.\end{align}
Here, the sum over $\kappa$ runs over all relevant partonic channels $\kappa = \{\kappa_a, \kappa_b; \kappa_1, \ldots,\kappa_N\}$ for the underlying $2\to N$ process (or $2\to N+L$ where $L$ denotes additional non-strongly-interacting final states). The $s_\kappa$ is the appropriate factor to take care of symmetry factors and flavor and spin averaging for each partonic channel. The $\df\Phi_N$ corresponds to the complete phase-space measure of the Born process with massless partons,
\begin{equation}  \label{eq:PhiN}
\int\!\df\Phi_N \equiv \frac{1}{2\Ecm^2} \int\!\frac{\df x_a}{x_a}\, \frac{\df x_b}{x_b}\, \int\! \df\Phi_N(q_a + q_b; q_1, \ldots, q_N, q)\,\frac{\df q^2}{2\pi}\, \df\Phi_L(q)  
\,,\end{equation}
where $\df\Phi_N(...)$ on the right-hand side denotes the standard Lorentz-invariant $N$-particle phase space, and $\df\Phi_L(q)$ the remaining nonhadronic phase space with total momentum $q$.  The variables appearing here and the restrictions we impose on the measurement function $X_N(\Phi_N)$ will be described further below.

Now consider $\widetilde \Tau_N$ in the exclusive $N$-jet limit $\widetilde \Tau_N\to 0$.  Since we are interested in the simplest form of the factorization theorem, we assume that the jets are well separated from each other and from the beams, with no strong hierarchies in the jet $p_T$s.  We also assume that if we are computing the cross section differential in $\widetilde \Tau_N^r$, we have parametrically $\widetilde \Tau_N^{\,r}\sim \widetilde \Tau_N^{\, r'}$.\footnote{This last assumption avoids the appearance of large nonglobal logarithms, $\ln(\widetilde \Tau_N^{\,r}/\widetilde\Tau_N^{\,r'})$. These logarithms will not appear when considering the cross section differential only in the total $\widetilde \Tau_N$.} For definiteness we assume that the components in the decomposition in \eq{linearmodes} below scale homogeneously, which will indeed be the case if the only $N$-jettiness that we measure is the total $\widetilde \Tau_N$.   In the exclusive $N$-jet limit, the final state consists of only soft radiation and so-called $n_r$-collinear energetic radiation which is collinear to one of the jet or beam directions $n_r$.    Here, the key property of $N$-jettiness is the presence of the minimum in its definition, which leads to a linear decomposition for both $\widetilde \Tau_N$ and  $\widetilde \Tau_N^{\,r}$. Namely, they can be decomposed as a sum of contributions coming from each of these types of emissions,
\begin{align}  \label{eq:linearmodes}
  \widetilde \Tau_N 
    &= \widetilde \Tau_N^{[n_a]} + 
    \widetilde \Tau_N^{[n_b]} + 
    \widetilde \Tau_N^{[n_1]} + \ldots + 
    \widetilde \Tau_N^{[n_N]} + 
    \widetilde \Tau_N^{[\rm soft]} 
    \,,
 & \widetilde \Tau_N^{\,r} &= 
   \widetilde \Tau_N^{[n_r]} +  \widetilde \Tau_N^{\,r[\rm soft]} 
  \,,
\end{align}
where the $[n]$ superscripts refer to the contribution from emissions collinear to the $n$-direction, and $[\rm soft]$ to soft emissions. For definiteness, we let 
\begin{align}
  n_a &=(1,\hat z)
  \,, \qquad 
  n_b  =(1,-\hat z) 
 \,,
\end{align}
where $\hat z$ is the physical beam direction.

Equation~\eqref{eq:linearmodes} encodes the fact that for all of the measures in \Tab{tab:measures}, the $n_r$-collinear emissions only contribute to the measurement in the $r$-th region, while the soft radiation contributes to all regions and can itself be decomposed as in \eq{linearregions}. This linearity is the key property that allows deriving a factorization theorem which decomposes the exclusive $N$-jet cross section into a product of functions for each type of radiation.  The basic form of the $N$-jettiness factorization theorem is~\cite{Stewart:2010tn}
\begin{align} \label{eq:roughfact}
\frac{\df\sigma_\kappa (\Phi_N)}{\df \widetilde \Tau_N} =
  {\rm tr}\  \widehat H_N^\kappa \otimes B_{\kappa_a} \otimes B_{\kappa_b}\otimes J_{\kappa_1} \otimes 
  \ldots \otimes J_{\kappa_N} \otimes \widehat S_N^\kappa
  \,.
\end{align}
Here, $\widehat H_N^\kappa$ is a hard function, $B_{\kappa_a,\kappa_b}$ are beam functions, $J_{\kappa_A}$ is a jet function for the $A$-th jet region,  and $\widehat S_N^\kappa$ is a soft function. A description of the variables these objects depend on will be given below.
We note immediately that $\widehat H_N^\kappa$ depends directly on the full partonic channel $\kappa$, as it contains the process-specific matrix elements, while $\widehat S_N^\kappa$ depends on $\kappa$ only via the color representations. The $J_{\kappa_A}$ depend on whether $\kappa_A$ is a quark or gluon that initiates the jet, and $B_{\kappa_a,\kappa_b}$ each depend on the flavor of the initial-state partons $\kappa_a$ and $\kappa_b$ and the type of initial-state hadrons. The $\widehat H_N^\kappa$ and $\widehat S_N^\kappa$ are both matrices in the color space of $\kappa$ which are traced over in \eq{roughfact}.

The precise form of the convolutions in \eq{roughfact}, as well as the definitions of the beam, jet, and soft functions, depends on the choice of jet and beam measures used in the $N$-jettiness observable. On the other hand, the hard function is not affected by these choices.  So far, we have been using the observables $\widetilde \Tau_N^{r}$ without specifying the method of fixing the jet axes $n_r$.  The form of the convolutions will generically depend on the jet axes choice.  We discuss below the observables $\Tau_N^{r}$ obtained after the axes minimization in \eq{mincriteria}.  The factorization in \eq{roughfact} holds for any jet axes choice that is within ${\cal O}(\lambda)$ of the minimized jet axes, where the power counting parameter $\lambda$ is defined below.

\subsection{Categorizing Measures by Power Counting}

To determine the structure of the convolutions in \eq{roughfact}, it is first instructive to form categories for the measures in \Tab{tab:measures} that share common features in their convolution structure. In particular, we classify them by how they scale with the SCET power counting parameter $\lambda \ll 1$.  Below, we use a light-cone decomposition of the momenta based on the jet axis $n_A$ satisfying $n_A^2=0$ as well as the auxiliary vector $\bar n_A$ obeying ${\bar n}_A^2=0$ and $n_A\cdot \bar n_A=2$.

An $n_A$-collinear mode within the $A$-th jet has momentum scaling as $(n_A\cdot p_i, {\bar n}_A\cdot p_i, p_i^{n_A\perp}) \sim {\bar n}_A\cdot p_i \, (\lambda^2,1,\lambda)$. Here and below we use the label $\perp$ to refer to components perpendicular to the respective jet axis $\vec n_A$, while $T$ indicates transverse momentum with respect to the beam. Considering all the jet measures in \Tab{tab:measures}, those with $\beta=2$ have $\widetilde \Tau_N^{[n_A]} \sim \lambda^2$ (which includes the geometric measures), while those with $\beta=1$ have $\widetilde \Tau_N^{[n_A]} \sim \lambda$. Since the components in the decomposition in \eq{linearmodes} scale homogeneously, the scaling of the corresponding soft momenta $\widetilde \Tau_N^{(A)[{\rm soft}]}$ must be the same as those of the corresponding collinear emissions. The soft momenta scale homogeneously, independent of the jet directions, so $p_s^\mu \sim \lambda^2$ for $\beta=2$ and  $p_s^\mu\sim\lambda$ for $\beta=1$.  The $\beta=2$ situation is known as an \SCETa observable, while the $\beta=1$ case is referred to as an \SCETb observable.

Since the convolutions in \eq{roughfact} are always between observables with the same $\lambda$-scaling, we can classify the jet measures by whether they are in \SCETa or in \SCETb. A similar classification can also be made for the beam measures. For collinear emissions along either of the two beams, $1/(2\cosh y_i) \simeq e^{-|y_i|}$ up to power corrections.  All beam measures having this exponential rapidity dependence are in \SCETa, while those measures with just $p_{Ti}$ are in \SCETb.  Summarizing the scaling of the measures in \Tab{tab:measures}, we have:
\begin{align}
 & \text{\SCETa jets \& beams:} 
   & & \text{Geometric(-R), Modified Geometric(-R),} \nn
  \\*
 & & &\text{Conical Geometric ($\beta=\gamma=2$)};  
  \nn\\*[5pt]
 & \text{\SCETa jets \& \SCETb beams:} 
   & & \text{Conical ($\beta=2$), XCone Default}; 
  \nn\\*[5pt]
 & \text{\SCETb jets \& beams:} 
   & & \text{Conical ($\beta=1$), Recoil-Free Default}; 
  \nn\\*[5pt]
 & \text{\SCETb jets \& \SCETa beams:} 
   & & \text{Conical Geometric ($\beta = 1$, $\gamma = 2$)},
\end{align}
though we have not made use of the last example in this paper.

Equation~\eqref{eq:linearmodes} for $\widetilde \Tau_N^{\,r}$  implies that the factorization theorem will have one convolution for each region it is differential in. For \SCETa cases we have convolutions in ($n_A\cdot p$)-momenta between the beam/jet functions and the soft function.  In contrast, for \SCETb cases we have convolutions involving transverse or $\perp$-momenta between the beam/jet functions and the soft function.    The homogeneous scaling for the components of $N$-jettiness also requires $\widetilde \Tau_N^{[n_r]}\sim \widetilde \Tau_N^{[n_{r'}]}$, such that all of the soft function convolution variables are of the same order in the power counting. If all jets and beams are in either \SCETa or \SCETb, then that theory's ingredients can be used for the main components of the analysis. In the mixed case of \SCETa jets with \SCETb beams, the restriction on the radiation imposed by the measurement together with the power counting implies that the modes in the $A$-th jet can have parametrically larger $\perp$-momenta relative to their $n_A$ axis than the modes in the beam do relative to the beam axis, since $p_{n_A \perp}^i \sim \lambda \gg p_{n_{a,b}\perp}^i \sim \lambda^2$.

One can also derive factorization theorems for $N$-jettiness measures with generic $\beta$.  For any $\beta$ such that $\beta-1 \gg \lambda$ these measures fall in the \SCETa category, and they lead to $\beta$-dependent jet, beam, and soft functions.  This is analogous to the factorization theorems derived in $e^+e^-\to $ dijets for general angularities \cite{Berger:2003iw,Hornig:2009vb} and their recoil-free variants \cite{Larkoski:2014uqa}.\footnote{In the case of recoil-free angularities, there is a smooth interpolation between \SCETa and \SCETb as $\beta$ goes from 2 to 1 \cite{Larkoski:2014uqa}.}  For simplicity we will not discuss the general $\beta$ case here, but instead focus on the representative cases of $\beta=1,2$.  

\subsection{Impact of Axes Minimization}

In general, the jet axes $n_A$ need not align perfectly with the jet three-momenta $\vec p_A$, as long as the difference is $\mathcal{O}(\lambda)$.  That said, the structure of the factorization theorem will simplify if we align the $n_A$ axes within ${\cal O}(\lambda^2)$ of the jet direction.  For jets defined with XCone, this alignment happens automatically for any of the $\beta=2$ measures (including the XCone default), as explained near \eq{nicemin} and discussed previously in \Ref{Thaler:2011gf} (see also \Ref{Ellis:2001aa,Thaler:2015uja}). For this reason, we will focus the remainder of our discussion on jet measures in the \SCETa category, including the XCone default.  This minimization implies that we are now discussing the specific $N$-jettiness observable $\Tau_N$ rather than the generic $\widetilde \Tau_N$.

The alignment of $n_A$ with $\vec p_A$ means that the jet momentum has ${\cal O}(\lambda^2)$ perpendicular momentum relative to this axis.  For all the geometric jet measures the perpendicular momentum is actually zero, and the component observables $\Tau_N^{A}$ then have a simple physical interpretation, since they measure the jet mass $m_A^2$ for each jet region via $\Tau_N^A = m_{A}^2/Q_A$ with $Q_A=2\rho E_A$~\cite{Jouttenus:2013hs}.  For our XCone default measure the perpendicular momentum is ${\cal O}(\lambda^2)$, however this same physical interpretation still applies, with the only difference being that $Q_A= R^2 E_A/ \cosh y_A$. On the other hand, for the conical geometric measure with $\beta=\gamma=2$ there is not a precise relation between $\Tau_N^A$ and $m_A^2$, unless we were to adopt as an additional approximation $y_i\simeq y_A$.

Without aligning the jet axes and the jet three-momenta, the jet functions in the $N$-jettiness factorization theorem would depend on both $Q_A \widetilde \Tau_N^{[n_A]}$ and the total $p_{n_A\perp}$, such as in the jet function
\be
J_{\kappa_A}\big(Q_A \widetilde \Tau_N^{[n_A]} - \vec p_{n_A\perp}^{\ 2},\mu\big).
\ee
Here, the two terms in $J_{\kappa_A}$ are both $\mathcal{O}(\lambda^2)$, and $\kappa_A$ indicates a quark or gluon.  With the axes minimization, the dependence on the transverse momentum drops out, and this becomes simply
\be
\label{eq:inclusivejetfunc}
J_{\kappa_A}\big(Q_A \Tau_N^{[n_A]},\mu\big).
\ee
These jet functions, which appear in the $N$-jettiness factorization theorem, are inclusive because the collinear radiation is always completely contained in the corresponding jet region. This means that they are a function of a single variable and do not depend on the jet boundary. However, the type of inclusive jet function we have does still depend on the jet measure. For instance, the geometric measures yield the standard inclusive hemisphere jet function, but we obtain a different inclusive jet function for the $\beta=\gamma=2$ conical geometric measure.

\subsection{Hadronic and Partonic Momentum Conservation}

The remaining ingredients that influence the form of the factorization theorem are momentum conservation and the choice of measurements $X_N$ made on the jets and the nonhadronic particles.  We will discuss the first issue here, before explaining why they impact the structure of the factorization theorem in the next subsection.

Momentum conservation says that
\begin{align} \label{eq:momcons}
 p_{\rm beam}^\mu =  p_a^\mu + p_b^\mu = q^\mu + \sum_{A} p_A^\mu \,,
\end{align}
where $p_A^\mu$ is the sum of all four-momenta for particles in region $A$, the $p_{a,b (\rm beam)}^\mu$ include the incoming proton momentum (momenta) minus the sum of the outgoing momentum of particles in the associated beam region, and $q^\mu$ is the total outgoing momentum of any nonhadronic particles. Even if the $N$-jettiness measurement specifies only a single beam region, we can divide the beam region in two by making an artificial split at zero rapidity into regions $a$ and $b$. This split is useful for the discussion below, since it makes it simpler to talk about the two beam functions that are important for the dynamics of the beam region.  We set $q^\mu=0$ for cases where the final state does not involve nonhadronic particles.

The largest ${\cal O}(\lambda^0)$ momentum  component from each jet and beam region in \eq{momcons} can be extracted by projecting along the associated $N$-jettiness axis,
\begin{align}  \label{eq:hardcomps}
  p_r^\mu = \omega_r \frac{n_r^\mu}{2} + {\cal O}(\lambda) \, .
\end{align}
This determines the variables appearing in the hard function
\begin{align} \label{eq:Hfunction}
  \widehat H_N
   = \widehat H_{N}(\{\omega_r n_r\},q,\mu)
 \,,
\end{align}
where $r$ runs over $a$, $b$, $1$, $\ldots$, $N$ in the set of variables in $\{\cdots\}$.\footnote{To emphasize that $\widehat H_N$ can always be written in terms of Lorentz-invariant phase-space variables, one can rewrite this as $\widehat H_N( \{ \omega_r \omega_{r'}\, n_r\cdot n_{r'}\},\, \{\omega_r\, n_r\cdot q\}, q^2)$ with $r$ and $r'$ running over $a$, $b$, $1$, $\ldots$, $N$.}  These phase-space variables include things like the transverse momentum $p_{T}^A$ and rapidity $\eta_A$ of each jet, as well as the overall rapidity of all non-forward radiation $Y$ which determines the boost of the partonic hard collision relative to the  center-of-mass frame.   These hard-function variables form the basis for the measurements we make on the jets as specified by $X_N(\Phi_N)$ in \eq{dsigmasingfull} where $q_r = \omega_r n_r/2$. The variables are not all independent, since momentum conservation correlates the large ${\cal O}(\lambda^0)$ components of \eq{momcons}.  This is the same as imposing momentum conservation for the underlying hard partonic process with incoming and outgoing massless partons,
\begin{align} \label{eq:hardmomcons0}
  \omega_a\, \frac{n_a^\mu}{2} + \omega_b\, \frac{n_b^\mu }{2}
     = q^\mu + \sum_A \omega_A\, \frac{n_A^\mu}{2}
   \,.
\end{align}
In particular, this formula is used to compute $\widehat H_N$ when integrating out hard modes by matching QCD to SCET using calculations of S-matrix elements in the two theories. And this momentum conservation appears above in $\df\Phi_N$ in \eq{PhiN}.  The same hard function in \eq{Hfunction} appears in the factorization theorem for exclusive jet cross sections for all choices of the $N$-jettiness jet and beam measures.

In \eq{hardmomcons0}, the beam variables can be rewritten in terms of the total center-of-mass energy $E_{\rm cm}$ and momentum fractions $x_{a,b}$ for the colliding partons in the hard collision via $\omega_a = x_a E_{\rm cm}$ and $\omega_b = x_b E_{\rm cm}$. The jet variables $\omega_A$ are chosen so that $\omega_A= 2 E_A + {\cal O}(\lambda)$, where $E_A$ is the true jet energy, and the presence of ${\cal O}(\lambda)$ contributions in this relation ensure that \eq{hardmomcons0} is exactly satisfied. The presence of these ${\cal O}(\lambda)$ terms does not affect the evaluation of the hard function in \eq{Hfunction}, where we may simply replace $\omega_A\to 2E_A$. This same replacement should be made in the formulas for the $Q_A$ factors appearing in the jet functions, which are otherwise given by the results in \Tab{tab:QA}.
However, the ${\cal O}(\lambda)$ terms can have implications for the convolutions between the jet, beam, and soft functions. To see explicitly how these ${\cal O}(\lambda)$ terms arise, it is convenient to project \eq{hardmomcons0} both along and transverse to the beam axis, giving
\begin{align} \label{eq:hardmomcons}
  \omega_a
   &= n_b\cdot q + \sum_A \omega_A\, \frac{n_b\cdot n_A}{2}
   \,,
  & & \omega_b  
   = n_a\cdot q + \sum_A \omega_A\, \frac{n_a\cdot n_A}{2}
   \,,  \\*
\label{eq:hardmomconsT}
 0 & = 2\, q_{T}^\mu + \sum_A \omega_A\, n_{AT}^\mu    
   \,.
\end{align}
The two equalities in \eq{hardmomcons} simply fix $\omega_{a,b}$  regardless of how precisely we specify the jet axes $n_A$, the jet variables $\omega_A$, or $q^\mu$.  This leaves the two constraints from \eq{hardmomconsT}, which will be very important in the next subsection.  These constraints involve $n_{AT}^\mu$, which is determined by the azimuthal angle $\phi_{n_A}$ for the axis of each jet region, but they do not depend on the longitudinal (rapidity) component of $n_A$.   

\begin{table}
\begin{tabular}{ccccc}
\hline \hline
 & (Modified) Geometric & Geometric-$R$ & Modified Geometric-$R$ & XCone Default \\
 \hline
  $Q_A=$  & $\rho_0\, \omega_A$
  & $\rho(R,y_A)\,\omega_A$ & $\rho_C(R,y_A)\,\omega_A$
  & $\dfrac{R^2}{2\cosh y_A}\,  \omega_A$\\ [5pt]
 \hline\hline
\end{tabular}
\caption{\label{tab:QA} Values of $Q_A$ for various measures. The approximation $\omega_A=2E_A$ is valid as long as the same replacement is made in the hard function.}
\end{table}

\subsection{Convolutions from Transverse Momentum Recoil}
\label{subsec:convolve}

We now show how the two constraints in \eq{hardmomconsT} can influence the form of the convolutions appearing in the factorization theorem.  Throughout this discussion, we assume that the jet axes $n_A$ and jet three-momenta $\vec p_A$ are perfectly aligned, as is the case for the $\beta =2$ measures with the minimized $ \Tau_N$.  We start with pure \SCETa observables before mentioning what happens with \SCETb beam measures.

To begin, imagine making highly granular measurements of the jet energies and directions with very fine $p_T^A$, $\eta_A$, and $\phi_A$ bins, as well as fully measuring the nonhadronic $q^\mu$.  In this situation, we have effectively completely measured the transverse vector $q_{T}^\mu$, the jet energies $E_A$, and the vectors $n_A$, so we actually have a measurement that is sensitive to the ${\cal O}(\lambda)$ amount by which the $\omega_A$ variables differ from $2E_A$. Here, the $A$-th jet's momentum can be written as
\begin{align}  \label{eq:pA}
 p_A^\mu
 = (2E_A- n\cdot p_A) \frac{n_A^\mu }{2} + n\cdot p_A\, \frac{\bar n_A^\mu}{2} + p_{A\perp}^\mu
   \,,
\end{align} 
where $n\cdot p_A \sim \Tau_N^A\sim \lambda^2$. We can therefore see that the components beyond $E_A n_A^\mu$ are ${\cal O}(\lambda^2)$ and do not have ${\cal O}(\lambda)$ projections on the axis transverse to the beam. If we consider transverse momentum conservation using the original momentum conservation in \eq{momcons}, and insert \eq{pA}, then we 
find that the balance of transverse momenta at ${\cal O}(\lambda)$ is given by
\begin{align} \label{eq:krTdefn}
  k_T^\mu &
  \equiv  p_{aT}^\mu + p_{bT}^\mu  
 =  q_T^\mu + \sum_A E_A\, n_{AT}^\mu 
  \,.
\end{align}
Using \eq{hardmomconsT} we can see that this is a small momentum $k_T^\mu\sim \lambda$. For the beam variables $p_{a,b}^\mu$, these ${\cal O}(\lambda)$ transverse components come from the   transverse momenta of radiation emitted in the beam regions (since the transverse momenta in the proton are $\sim\Lambda_{\rm QCD}$ which is much smaller). For the jet components, this ${\cal O}(\lambda)$ momentum comes from the mismatch between $\omega_A$ and $2 E_A$, which we can see explicitly by using \eq{pA} in \eq{hardmomconsT} to give
\begin{align}
  k_T^\mu &
 = \sum_A \Big(E_A - \frac{\omega_A}{2}\Big) n_{AT}^\mu  
  \,.
\end{align}

With the assumptions above, the constraint in \eq{krTdefn} is present because by making such a granular measurement, we have indirectly measured $k_T^\mu$, and hence the total transverse momentum recoil of the beam radiation.  This measurement therefore leads to $p_T$-dependent beam functions in the factorization theorem, which appear as
\begin{align} \label{eq:beampTconv}
   \int \df^2p_{T}\:  B_{\kappa_a}(t_a, x_a,  \vec p_{T},\mu)\:
       B_{\kappa_b}(t_b, x_b,  \vec k_T - \vec p_{T},\mu) \,.
\end{align}
Here $t_a=\omega_a \Tau_N^{[n_a]}$ and $t_b=\omega_b \Tau_N^{[n_b]}$ involve the variables that are convolved with the soft function.  The double differential beam functions $B_{\kappa_a}(t_a, x_a,  \vec p_{T},\mu)$ were discussed in \Refs{Jain:2011iu,Mantry:2009qz}.  In \Ref{Kang:2013nha}, examples where transverse momentum convolutions connect a jet and beam function were discussed for an \SCETa type $1$-jettiness in deep inelastic scattering, and \eq{beampTconv} is the analog of the center-of-mass $1$-jettiness variable considered there, except with the jet function replaced by a second beam function.  The double differential factorization theorem with an explicit measurement of $0$-jettiness and $k_T$ in \SCETa was derived in \Ref{Procura:2014cba}, and involves precisely the combination in \eq{beampTconv}.

To obtain a simpler factorization theorem that does not involve $p_T$-dependent beam functions, we just have to perform a less granular measurement that does not constrain every aspect of the final state.  For cases with external nonhadronic particles, the simplest approach is to not fully constrain all components of $q_T^\mu$, for example by specifying $q_T$ only within a bin centered on $q_T^{\rm central}$ with width $> \lambda q_T^{\rm central}$. Since $\lambda\simeq m_A/E_A\simeq 0.1$, this corresponds to the typical size of bins that are already used in experimental analyses (unless they are only interested in measuring $q_T$). This method was used in \Ref{Stewart:2009yx} when deriving the active-parton factorization theorem for beam thrust or $0$-jettiness, where $q_T$ was simply not measured.  For beam thrust there are no jets, so $q_T^{\rm central}=0$, but this approach works equally well for $(N\ge 1)$-jettiness where $q_T^{\rm central}\sim \lambda^0$ is large.  Once one uses this coarser $q_T$ binning, there are no other ${\cal O}(\lambda)$ constraints on the transverse momenta.  In particular, specifying the bin for $q_T$ yields an additional \emph{unrestricted} integration over $k_T^\mu$ which appears in \eq{beampTconv} when deriving the factorization theorem. Therefore, we obtain independent transverse integrals over the two beam functions, $\int \df^2 p_T B_{\kappa}(t,x,\vec p_{T},\mu) = B_{\kappa}(t,x,\mu)$, and only these $p_T$-independent beam functions appear in the $N$-jettiness factorization theorem, as in
\begin{align}  \label{eq:BBincl}
  B_{\kappa_a}(t_a, x_a, \mu)\:
       B_{\kappa_b}(t_b, x_b, \mu) \,.
\end{align}

Alternatively, for cases where $N\ge 2$, we can exploit the fact that we do not need to make finely-binned measurements of the jet energies or jet $p_T$s.  We can instead be satisfied with a measurement with center $p_T^{\rm central}$ in a bin of width $> \lambda p_T^{\rm central}$, which could be for example using a bin centered at $500\:{\rm GeV}$ with width $50\:{\rm GeV}$. This can be applied to both cases with ($q\ne 0$) or without ($q=0$) additional nonhadronic particles.  Since we can now vary by ${\cal O}(\lambda)$ at least two of the $\omega_A$ variables, we again loosen the constraint fixing $k_T^\mu$ and we can again freely integrate over this variable, and hence also obtain \eq{BBincl}.  Both of these approaches to obtaining the simpler form of beam functions in \eq{BBincl} require making less granular measurements when specifying $X_N$, but still remain fully sufficient for all standard LHC jet-style measurements. The only cases where \eq{beampTconv} become relevant is if we are actually interested in making a jet measurement so finely-binned that we can infer the small $p_T$ spectrum of the beam radiation.

Just like for the jet function in \eq{inclusivejetfunc}, the beam functions in \eq{BBincl} are inclusive because collinear radiation along the beam directions is completely contained in the beam regions.  Thus, they do not depend on the boundaries between the beam and jet regions. In principle, they could still depend on the beam measure, but because of \eq{mintoinv}, for all the \SCETa beam measures we consider here, they are always given by the standard inclusive hemisphere beam functions~\cite{Stewart:2009yx, Stewart:2010qs}.

It is interesting to consider how the above arguments change if we maintain \SCETa measures for the jets (and aligned jet axes obtained from minimization) but now consider a \SCETb measure for the beam; this is the case encountered in the XCone default measure.   In this situation, we still have inclusive jet functions that do not depend on $p_{n_A\perp}$ as in \Eq{eq:inclusivejetfunc}.  The key change is that now the $N$-jettiness measurement forces the beam transverse momenta to be smaller, $p_{aT}^\mu\sim p_{bT}^\mu\sim \lambda^2$, and the resulting \SCETb beam functions are of the broadening variety with $t_{a} = \Tau_N^{[n_{a}]}$ and $t_{b} = \Tau_N^{[n_{b}]}$ variables that are themselves ${\cal O}(\lambda^2)$.   In addition to the renormalization scale $\mu$, the beam functions depend on a rapidity renormalization scale $\nu$, in the combination $\nu/\omega_{a,b}$. The $\nu$ scale is needed to sum logarithms associated with rapidity divergences that appear from the separation of modes in the beam and soft functions~\cite{Chiu:2011qc,Chiu:2012ir}.   From \eq{krTdefn}, we must also have $k_T^\mu= p_{aT}^\mu+p_{bT}^\mu =q_T^\mu + \sum_A E_A n_{AT}^\mu \sim\lambda^2$. Once again we can integrate over $k_T^\mu$ either by considering a bin for $q_T^\mu$ or a bin for two of the jet energies $E_A$.  In this case, the bins need only have a size of $> \lambda^2 q_T^{\rm central}$ or $> \lambda^2 E_A^{\rm central}$ in order to sufficiently integrate over $k_T^\mu$ such that we get $p_T$-independent beam functions, as in \eq{BBincl}.

\subsection{Factorization Theorems for $N$-jettiness}

We now have all the ingredients needed to assemble the factorization theorem for $N$-jettiness for various jet and beam measures. For jet and beam measures in \SCETa, the mathematical derivation of this factorization theorem follows closely the detailed derivation given for beam thrust in \Ref{Stewart:2009yx}, or for DIS 1-jettiness in \Ref{Kang:2013nha}, which we therefore will not bother to repeat here. The $N$-jet case has also been discussed in some detail in \Refs{Jouttenus:2011wh, Gaunt:2015pea}. The required ingredients in the derivation have all been discussed in the previous subsections.

With jet and beam measures in the \SCETa category, axes determined by minimization, and the choice for $X_N$ that does not directly or indirectly measure the transverse momentum of the beam radiation, the factorization theorem in \eq{roughfact} becomes
\begin{align}  \label{eq:factthmscet1twobeams}
 &\frac{\df\sigma_\kappa(\Phi_N)}
  {\df\Tau_N^{a} \df\Tau_N^{b} 
    \df\Tau_N^1\dotsb \df\Tau_N^N}
  \\
  &\qquad
  =  {\rm tr}\: 
  \widehat H_N^{\kappa}(\{\omega_r n_r\},q,\mu)
  \int\! \Bigl[\prod_r \df\Tau_N^{[n_r]}\Bigr]
  \omega_a B_{\kappa_a}\bigl(\omega_a \Tau_N^{[n_a]},x_a,\mu\bigr)\,
  \omega_b B_{\kappa_b}\bigl(\omega_b \Tau_N^{[n_b]},x_b,\mu\bigr)
  \nn\\
 & \quad \qquad \times
  Q_1 J_{\kappa_1} \bigl(Q_1 \Tau_N^{[n_1]},\mu \bigr)
   \cdots Q_N J_{\kappa_N} \bigl(Q_N\,\Tau_N^{[n_N]},\mu\bigr)\,
  \widehat S_N^{\kappa} \Bigl( \bigl\{ \Tau_N^{r}-\Tau_N^{[n_r]} \bigr\},
  \Bigl\{ \frac{\omega_r n_r}{Q_r} \Bigr\},\mu \Bigr) 
 \,,\nn
\end{align}
where $r$ and $r' =a,b,1,\ldots,N$ and all of the convolutions are now made explicit.  Here, the soft function $\widehat S_N^\kappa$ depends on the $N+2$ observables $\Tau_N^r$. It is a scalar function of the variables $\{ \omega_r n_r^\mu/Q_r\}$, which encode the dependence on the angles between various beam and jet directions through their dot products. Although not indicated by our notation, the soft function also depends on the size and shape of the jet regions through the precise definition of the jet and beam measures used to define these observables.  Both the jet functions and beam functions in \eq{factthmscet1twobeams} are of the inclusive variety, and hence do not depend on the boundaries between the jet or beam regions.  The beam functions also contain the nonperturbative parton distributions $f_j(\xi,\mu)$ through a factorization from the perturbative radiation into calculable coefficients ${\cal I}_{ij}$~\cite{Fleming:2006cd, Stewart:2009yx, Stewart:2010qs},
\begin{align}  \label{eq:factB}
  B_{i}(\omega k,x,\mu)
    &= \sum_j \int_x^1\! \frac{\df z}{z}\,
    {\cal I}_{ij}(\omega k, z,\mu)\, f_j \Bigl(\frac{x}{z},\mu \Bigr)
\,.\end{align}
With geometric (and related) measures, \eq{factthmscet1twobeams} was the version of the $1$-jettiness factorization theorem used for the analysis in Ref.~\cite{Jouttenus:2013hs}.

For the various geometric measures, the $Q_A$ factors needed for \eq{factthmscet1twobeams} are given above in \Tab{tab:QA}. For the $\beta=\gamma=2$ conical geometric measure we let $Q_A=R^2\omega_A$.  For this measure, the inclusive jet functions become $J_{\kappa_A}(Q_A \Tau_N^{[n_A]},y_A,\mu)$ in \eq{factthmscet1onebeam}, due to the $\cosh y_A/\cosh y_i$ weighting factor in the jet measure. Thus, they are not just the standard hemisphere jet functions.    Similarly, for this case we also will have a soft function that can depend on the $y_A$ variables.  

In \eq{factthmscet1twobeams} we are differential in two beam regions, $\Tau_N^a$ and $\Tau_N^b$.  If we only want to consider a single beam region and measurement observable $\Tau_N^{\rm beam} = \Tau_N^a +\Tau_N^b$, then it is possible to simplify the form of the factorization theorem. Using the corresponding collinear projection,  $\Tau_N^{a[n_a]}+\Tau_N^{b[n_b]} = \Tau_N^{[n_{\rm beam}]}$, yields a ``double-beam function'' for \SCETa measures
\begin{align} \label{eq:BB}
 B\!B_{ij}(\omega_a \Tau_N^{[n_{\rm beam}]}, \omega_b \Tau_N^{[n_{\rm beam}]}, x_a, x_b,\mu)
   &= \omega_a\omega_b\!\! \int\! \df k\,  B_i\big(\omega_a k,x_a,\mu\big)\,
      B_j\big(\omega_b(\Tau_N^{[n_{\rm beam}]}-k),x_b,\mu\big)  \,.
\end{align}
Projecting the soft function in the same way, using $\Tau_N^{a[\rm soft]}+\Tau_N^{b[\rm soft]} = \Tau_N^{\rm beam [soft]}$, this reduces \eq{factthmscet1twobeams} to
\begin{align} \label{eq:factthmscet1onebeam}
 &\frac{\df\sigma_\kappa(\Phi_N)}
 {\df\Tau_N^{\rm beam} \df\Tau_N^1\dotsb \df\Tau_N^N} 
  \\*
 &\qquad
  = {\rm tr}\: 
  \widehat H_N^{\kappa}(\{\omega_r n_r\},q,\mu)
  \int\! \Bigl[\prod_r d\Tau_N^{[n_r]}\Bigr] 
  B\!B_{\kappa_a \kappa_b}\bigl(\omega_a \Tau_N^{[n_{\rm beam}]},\omega_b \Tau_N^{[n_{\rm beam}]}, x_a,x_b,\mu \bigr)
  \nn\\*
 &\quad\qquad
  \times Q_1 J_{\kappa_1}\bigl(Q_1\,\Tau_N^{[n_1]},\mu\bigr) 
   \cdots Q_N J_{\kappa_N}\bigl(Q_N\,\Tau_N^{[n_N]},\mu\bigr)\,
  \widehat S_N^{{\kappa}\rm (I)}\Bigl( \bigl\{ \Tau_N^{r}-\Tau_N^{[n_r]} \bigr\}, \Bigl\{ \frac{\omega_r n_r}{Q_r} \Bigr\},\mu\Bigr) 
  \,,\nn
\end{align}
where now $r={\rm beam},1,\ldots,N$. For the modified geometric(-R) measure, the soft function $\widehat S_N^{{\kappa}\rm (I)}$ in \eq{factthmscet1onebeam} has a $C$-parameter-type measurement for its $\Tau_N^{\rm beam}$ observable and thrust-type measurements for the jet observables $\Tau_N^{A}$, and \eq{factthmscet1onebeam} involves the standard inclusive hemisphere jet functions.  

Next, we consider the mixed measure case, with \SCETa jet measures and \SCETb beam measures, still with jet axes determined by minimization and a choice of $X_N$ that is insensitive to transverse momentum of the beam radiation. For this case, there has not yet been any literature providing a detailed mathematical derivation of a factorization theorem. Factorization theorems have been worked out for pure \SCETb measurements of event shapes in $e^+e^-\to $ dijets~\cite{Chiu:2011qc, Becher:2011pf, Chiu:2012ir, Becher:2012qc, Larkoski:2014uqa}, and active-parton factorization theorems have also been derived for $pp \to H$ with an $E_T$ jet veto~\cite{Tackmann:2012bt} or $p_T^{\rm jet}$ veto~\cite{Becher:2012qa, Tackmann:2012bt, Banfi:2012jm, Becher:2013xia, Stewart:2013faa}; see also~\cite{Becher:2015gsa} for transverse thrust.  Experience from these results enables us to anticipate the form of the convolutions that will appear between the beam and soft functions in the mixed measure $N$-jettiness case. So even though the complete derivation of the factorization theorem for this case is beyond the scope of this work, we can still put the information collected above together to anticipate its structure.

For \SCETb beam measures we expect the double-beam function to be given by
\begin{align} \label{eq:BB2}
 B\!B_{ij}\Big(k, x_a, x_b, \mu,\frac{\nu}{\omega_a},\frac{\nu}{\omega_b}\Big)
   &= \int\! \df k'\,  B_i\Big(k',x_a, \mu, \frac{\nu}{\omega_a}\Big)\,
      B_j\Big(k-k', x_b, \mu,\frac{\nu}{\omega_b}\Big)  \,.
\end{align}
The individual beam functions here are of the broadening type and involve the rapidity scale parameter $\nu$~\cite{Chiu:2012ir}.  For \SCETa jet measures and a single \SCETb beam measure we then expect a factorization theorem of the form
\begin{align} \label{eq:factthmscet12onebeam}
 &\frac{\df\sigma_\kappa(\Phi_N)}
 {\df\Tau_N^{\rm beam} \df\Tau_N^1\dotsb \df\Tau_N^N} 
  \\*
 &\qquad
 =  {\rm tr}\, 
  \widehat H_N^{\kappa}(\{\omega_r n_r\},q,\mu)
  \int\! \Bigl[\prod_r \df\Tau_N^{[n_r]}\Bigr] 
  B\!B_{\kappa_a \kappa_b}\Bigl(\Tau_N^{[n_{\rm beam}]}, x_a,x_b, \mu,\frac{\nu}{\omega_a},\frac{\nu}{\omega_b}\Bigr)
  \nn\\
 &\qquad \quad
  \times Q_1 J_{\kappa_1}\bigl(Q_1\,\Tau_N^{[n_1]},\mu\bigr)
   \cdots Q_N J_{\kappa_N}\bigl(Q_N\,\Tau_N^{[n_N]},\mu\bigr)\,
  \widehat S_N^{{\kappa}\rm (I/II)}\Bigl( \bigl\{ \Tau_N^{r}-\Tau_N^{[n_r]} \bigr\},
  \Bigl\{ \frac{\omega_r n_r}{Q_r} \Bigr\} ,\mu,\frac{\nu}{\mu} \Bigr) 
  \,.\nn
\end{align}
This is the factorization formula that is relevant for the XCone default measure, with the $Q_A$ factors given above in \Tab{tab:QA}. Note that here the soft function $\widehat S_N^{{\kappa}\rm (I/II)}$ has broadening-type variables convolved with the beam functions, and has dependence on the scale $\nu$ which compensates the $\nu$ dependence in the double-beam function. The conical measure with $\beta = 2$ will have an analogous factorization theorem but requires different jet and soft functions that take into account that the jet measure cannot be written as $n\cdot p_i\, \omega_A/Q_A $ with some $Q_A$. We leave a detailed mathematical analysis and proof of the active-parton factorization theorem in \eq{factthmscet12onebeam} to future work.   It will also be interesting to test it against fixed-order predictions for these $N$-jettiness distributions.

The other main class of measures in \Tab{tab:measures} are those that have both jet and beam measures in the \SCETb category. This includes the recoil-free default XCone measure, as well as the conical measure with $\beta = 1$. Once again there has not yet been a detailed mathematical analysis of this case in the literature, but from our previous analysis and from experience with simpler cases, we can anticipate the form of the associated factorization theorem.  With axes determined by minimization, and with a choice of $X_N$ that is again insensitive to the total transverse momentum of the beam radiation, we expect the appropriate factorization theorem to contain the same \SCETb double-beam function in \eq{BB2} with no additional recoil convolutions. This would be analogous to the factorization theorem for recoil-free broadening in $e^+e^-$ collisions in \Ref{Larkoski:2014uqa}.
In contrast to \eq{factthmscet12onebeam}, the jet functions must now be of the broadening type and likely also depend on a rapidity scale $\nu$. The corresponding soft function $S_N^{{\kappa}\rm (II)}$ now only depends on convolution variables of the broadening type and has to cancel the $\nu$ dependence of both beam and jet functions.  We again leave a detailed mathematical analysis and proof of the active-parton factorization theorem for this case to future work. 

\section{Conclusions}
\label{sec:conclude}

In this paper, we introduced the new XCone jet algorithm, which is based on the $N$-jettiness event shape. XCone is an exclusive cone algorithm that finds a fixed predefined number of jets. Exploiting the measure flexibility inherent to $N$-jettiness, we defined a new conical geometric measure that combines the geometric measure, which is theoretically motivated and preferred, with the conical measure, which has already been proven to be experimentally robust in the context of jet substructure techniques using $N$-subjettiness. In a companion paper \cite{Thaler:2015xaa}, we present three physics case studies to highlight how XCone can be beneficial to a variety of LHC analyses. In particular, XCone is capable of resolving overlapping jets without requiring a separate split/merge step, and allows for a continuous transition from the resolved regime of well separated jets to the boosted regime of overlapping jets.

Our focus in this paper was on the case $\gamma = 1$, for which the beam measure scales as $p_T$, such that $\Tau_N$ minimization is roughly the same as minimizing the total unclustered $p_T$.  By changing $\gamma$, one changes whether jets are found preferentially in the central or forward parts of the detector.  In the future, it would be interesting to study the impact and utility of different $\gamma$ values, especially $\gamma = 2$ which is the natural value from the original geometric measure.  At present, the XCone code is limited to $\gamma = 1$, primarily because our method to find seed axes employs the existing longitudinally-invariant generalized $k_T$ algorithm.  It is possible to build recursive clustering algorithms optimized to find seed axes for any given $\Tau_N$ measure, which is planned for future work.

In constructing the XCone algorithm, we have chosen a specific measure for both the $N$-jettiness partitioning into jet and beam regions as well as the jet axis finding via the overall $N$-jettiness minimization.  This has lead to an interesting compromise, where in order for the XCone default measure to use dot-product distances in the jet partitioning, the jet regions could not be perfectly stable cones (meaning the jet axis is not exactly aligned with the total jet momentum).   One could imagine loosening the requirement of $\Tau_N$ minimization, though, to define an array of exclusive jet algorithms.  Following the idea that jet axis finding and jet region finding can be regarded as two distinct steps, one could use any exclusive clustering algorithm to find jet axes and only use $\Tau_N$ for defining the jet partitions.  Alternatively, if one wants the jet axis to be perfectly aligned with the jet momentum, one could build an exclusive cone jet algorithm that directly searches for $N$ mutually stable perfect cones.  More generally, it is worth reexamining the potential of exclusive jet algorithms at hadron colliders, and XCone provides a clear proof of concept with interesting physics applications \cite{Thaler:2015xaa}.

Beyond just being an exclusive jet algorithm that finds a fixed number of jets, XCone can be adapted to become an inclusive jet algorithm  that finds a variable number of jets by analyzing the distribution of $\Tau_N$ for different $N$.  For an event with $M$ jets, $\Tau_N$ should be large when $N < M$ and small when $N \geq M$, producing a sharp downward transition in the value of $\Tau_N$ when $N = M$.  Therefore, one could iteratively increase the value of $N$ until $\Tau_N$ undergoes this transition, either by measuring the ``slope'' $\df \Tau_N / \df N$ or by imposing a fixed $\Tau_{\rm cut}$.\footnote{Because XCone only finds a local minimum by default, there is no guarantee that $\Tau_N$ is a strictly decreasing function of $N$, though in practice this is a small effect when using the heuristic in \Eq{eq:pdeltaheuristic}.}  Using XCone as an inclusive jet algorithm could potentially be useful for jet counting in event samples with a variable number of jets, for accurate event reconstruction in the face of hard initial state radiation, or for improving background discrimination by dividing an event sample into exclusive $N$-jet bins.

Finally, we anticipate that the XCone default measure will be used in future $N$-jettiness theoretical calculations.  Since XCone is IRC safe, there are no obstacles for performing fixed-order or resummed calculations for any of the measures studied here.  While jet and beam measures that are linear in the particle momenta (like the XCone default measure) are simplest when using factorization to carry out calculations, the discussion in \Sec{sec:factorization} implies that the same SCET-based methods can also be applied for other measures.  Ultimately, we look forward to comparing precision XCone-based calculations to precision XCone-based measurements at the LHC.

\acknowledgments{
We thank Daniele Bertolini, Matteo Cacciari, Steve Ellis, Duff Neill, Gavin Salam, Gregory Soyez, Wouter Waalewijn, and Ken Van Tilburg for helpful conversations.  This work was supported by the Offices of Nuclear and Particle Physics of the U.S. Department of Energy (DOE) under Contracts DE-SC00012567 and DE-SC0011090.  I.S. is also supported by the Simons Foundation Investigator grant 327942.  F.T. is also supported by the DFG Emmy-Noether Grant No.~TA 867/1-1.  J.T.\ is also supported by the DOE Early Career research program DE-SC0006389 and by a Sloan Research Fellowship from the Alfred P.\ Sloan Foundation.  C.V.\ is also supported by the U.S.~National Science Foundation under Grant Nos.~NSF-PHY-0705682, NSF-PHY-0969510 (LHC Theory Initiative).  T.W.\ is also supported by the MIT Undergraduate Research Opportunities Program (UROP) through the Paul E. Gray Endowed Fund.}

\bibliographystyle{JHEP}
\bibliography{njettiness}
\end{document}